\documentclass[useAMS,usenatbib]{mn2e}

\usepackage{amsmath,amssymb}
\usepackage{amsfonts}
\usepackage{bbold}
\usepackage{graphicx}
\usepackage{color}
\usepackage[]{hyperref}
\usepackage{transparent}
\usepackage{rotating}
\usepackage{caption}
\usepackage{subcaption}
\usepackage{natbib}
\usepackage{color}
\usepackage[normalem]{ulem}

\usepackage{xspace}

\def\aap{{A\&A}}
\def\mnras{{MNRAS}}
\def\apj{{ApJ}}
\def\apjl{{ApJL}}

\newcommand{\PV}{ }
\newcommand{\FC}{ }

\title[RWI near  Schwarzschild  BH]{Impact of Schwarzschild black hole's gravity upon the Rossby wave instability}
\author[F. Casse, P. Varniere \& Z. Meliani]{F. Casse$^{1}$\thanks{\href{mailto:fcasse@apc.univ-paris7.fr}{fcasse@apc.univ-paris7.fr}}, P. Varniere$^{1}$ \& Z. Meliani$^{2}$\\
$^{1}$APC - Sorbonne Paris Cit\'e, Universit\'e Paris Diderot, CNRS/IN2P3, CEA/Irfu, Observatoire de Paris\\
10, rue Alice Domon et L\'eonie Duquet, F-75205 Paris Cedex 13, France\\
$^{2}$LUTh - Laboratoire de l'Univers et de ses Th\'eories,- CNRS UMR 8102,  Observatoire de Paris, Universit\'e Paris Diderot,\\
5 place Jules Janssen 92195 Meudon Cedex, France} 

\begin{document}

\date{Accepted YEAR MONTH DAY. Received YEAR MONTH DAY; in original form YEAR MONTH DAY}

\pagerange{\pageref{firstpage}--\pageref{lastpage}} \pubyear{YEAR}

\maketitle

\label{firstpage}

\begin{abstract}
\FC{In an early work}
the Rossby Wave Instability (RWI) has been proposed to explain variability thought to originate in the close vicinity of black-holes 
but this was done in the pseudo-Newtonian approach. Here we present the first general relativistic hydrodynamics 
simulations of this instability not only proving its theorized existence in a full general relativistic (GR) environment but also studying the effect of  the strong gravity on the instability.
To that end we performed a set of simulations increasingly closer to the black hole with our new GR version of the MPI-AMRVAC code.
 This allows us to study the minute changes in the behaviour of the instability. 
 We found that the pseudo-Newtonian approach gives adequate results provided that time shifting induced by the black hole gravity is taken into account. Hence, to view the disc as a distant observer would a full GR ray-tracing post treatment of the simulations is a must.

\end{abstract}

\section{Introduction}

        The Rossby-Wave Instability (RWI) was first introduced in galactic disk by \citet{Lovelace78} and dubbed, at the time, the negative-mass instability.
	The difficulty of finding a physical setup fulfilling its criterion made it rarely studied outside of carefully crafted configuration \citep{Lovelace99,Li00,Li01}.	
 	Over the past decade the RWI has been the subject of an increasing interest from a wide range of astrophysical communities
	 ranging from  planet-formation \citep{VarnT06,Meheut10,Lyra12, Lin12}  
	 to fast variability of microquasars \citep{TV06,VTR11,VTR12,Vin13} or the flares of Sgr A$^\star$ \citep{Tagger06,Vin14}.
	 That last case comes from the fact that circularized matter falling toward a black hole will meet the criteria for the RWI. 
	Such phenomena has the propensity to
	create a wide variety of light-curves displaying timescales coherent with the observed radiative emission coming from the galactic centre.
	
	Here we will be looking into more details at how the location of the Rossby vortices with respect to the last stable orbit 
	of the black hole does influence its physics and what 
	changes,  if any, we should be looking for in observations. To that end we will run a set of five GR hydrodynamic simulations,
	each fulfilling the RWI criterion at a different locations in the disc.
	We want to stress here that the simulations of our set are independent from one another.   Indeed, considering in a single simulation an 
	 annulus of gas falling onto the black hole would prevent us from  disentangling general relativistic effects from dynamical 
	effects stemming from the motion of the gas.
	\\
	In the next section we will briefly present the basic features of the RWI and the general relativistic hydrodynamics numerical 
	code MPI-AMRVAC used to perform the GRHD simulations presented in this paper. 
	In Sect.\ref{sec:setup}, we will describe the initial setup of the various simulations as well as the diagnostics to 
	track the changes in the behaviour of the instability. 
	We also present in this section the overall methodology. 
	Sect.\ref{sec:towardLSO} is devoted to the presentation of our 2D simulations and the complete display of the RWI features 
	depending on the location of 
	the instability with respect to the central black hole. 
	In Sect.\ref{Sec:3Dsimus} we describe the 3D GRHD simulations of the RWI and their comparison to the previous 2D computations. 
	Finally, Sect.\ref{Sec:conc} will present our conclusions.


\section{RWI in GRHD}
\label{sec:RWI_diag}
        As we are interested in tracking the changes in the evolution of the RWI as the instability develops closer and closer to 
        a non-rotating black hole, we will briefly summarize its basic features in the Newtonian case as well as the code we will be using.
       Our presentation of the RWI is by no mean exhaustive and for a proper review see for example \citet{Lovelace14}.

\subsection{Basic features of the RWI}

        The Rossby Wave Instability is a global instability found in differentially rotating disk exhibiting an extremum of the vortensity ${\cal L}$, whose definition for
        unmagnetized 2D disk is 
\begin{eqnarray}        
        {\cal L} = (\nabla \times \text{\bf v})_\perp /\Sigma  \label{eq:critere}
\end{eqnarray}              
         where $\bf{v}$ is the velocity of the disc matter while $\Sigma$ is its height-integrated density ($\perp$ stands for the component perpendicular the the disc plane).  
         The position of the extremum is called the corotation radius $r_c$ of the wave, meaning that at this radius the wave and the gas have the same velocity. 

        For a long time the existence of such extremum was a bottleneck that made this instability underused.  
        Since then, a few scenarios creating \lq bumps\rq\  in the vorticity have been found leading to more in-depth study of the RWI, especially  in the case of protoplanetary disk.

        Once there is an extremum of ${\cal L}$ in the disk, a standing wave pattern can form and creates vortices.
        As the energy and angular momentum is transferred from inside the corotation (positive energy) to outside the corotation we get an exponential growth of the vortices corresponding to the linear growth stage of the instability. Then, either the \lq bumps\rq\  is eaten away leading to a decrease of the instability which then dies with the extremum, 
        or it is replenish (by whatever caused it in the first place) and we get in a saturation phase. 
       
\subsection{General relativistic formalism in MPI-AMRVAC}
  
As we intend to study the evolution of the RWI in the close vicinity of a Schwarzschild black hole it is mandatory to use a numerical code encompassing a full general relativistic hydrodynamic framework. 
Strong of our experience with the existing MPI-AMRVAC code \citep{Holst12}, we have implemented a full GR support in it. Here we briefly present our approach and 
a test of the code. 

\subsubsection{GRHD equations of polytropic fluids}

      The governing equations of GRHD stem from fluid mass, momentum and energy conservations expressed in a general relativistic framework. The perfect non-magnetized fluid is  
       characterized by its momentum $4-$vector $\rho u^\alpha$ and stress-energy tensor  $T^{\mu\nu}$, namely
\begin{equation}
T^{\mu\nu}=\rho hu^\mu u^\nu + Pg^{\mu\nu}\nonumber
\label{Eq:stressenergy}
\end{equation}
where $\rho$ is the proper density of the fluid, $u^\mu$ is the contravariant $4-$velocity, $P$ the thermal pressure and $h$ the specific enthalpy of the fluid. The gravitational influence 
of the central object appears through the presence of a non-flat metric $g^{\mu\nu}$ whose expression depends on the nature of the central object. The general expression of a line element is, 
using a $(3+1)$ splitting of spacetime
\begin{equation} 
ds^2= -\alpha^2c^2dt^2 + \gamma_{ij}(dx^i+\beta^idt)(dx^j+\beta^jdt)\nonumber
\label{Eq:lineelem}
\end{equation}
where $\gamma_{ij}$ is the spatial metric tensor while $\alpha$ is the lapse function and $\beta^i$ is the shift vector. Let us note that latin indices stand for spatial coordinates while greek indices are linked to all spacetime coordinates.
In this first study we will consider non-rotating black holes described in Boyer-Lindquist coordinates $(r,\theta,\varphi)$ by the Schwarzschild metric, namely
\begin{equation}
\alpha^2=\left(1-\frac{2r_g}{r}\right)=\frac{1}{\gamma_{rr}} ,  \gamma_{\theta\theta}=r^2Ê ,Ê\gamma_{\varphi\varphi}=r^2\sin\theta\nonumber
\label{Eq:BHmetric}
\end{equation}
where $r_g=GM/c^2$ is the gravitational radius of the black hole. The temporal coordinate $t$ used in our description is consistent with time experienced by a so-called Newtonian observer located at infinity where the metric tensor converges towards a flat spherical metric expression.
Mass and momentum conservations are expressed as the vanishing covariant derivative of the fluid momentum $4-$vector and stress-energy tensor 
\begin{equation}
\rho u^\mu_{;\mu} = 0 \ ; \ T^{\mu\nu}_{;\mu} = 0\nonumber
\label{Eq:MAsscons}
\end{equation}
where $;\mu$ stands for the covariant derivative with respect to coordinate $\mu$.  The corresponding set of GRHD equations is
 \begin{eqnarray}
 \partial_t (\sqrt{\gamma}D) +\partial_j\left(\sqrt{\gamma}D\left(\alpha\text{v}^j-\beta^j\right)\right) = &0&\nonumber\\
 \partial_t (\sqrt{\gamma}S_i) + \partial_j\left(\sqrt{\gamma}\left[S_i(\alpha\text{v}^j-\beta^j)+\alpha P\delta_i^j\right]\right) = &&\nonumber\\
 \sqrt{\gamma}\left\{-(\xi-P)\partial_i\alpha + \frac{\alpha}{2}\left(S^j\text{v}^k+P\gamma^{jk}\right)\partial_i\gamma_{jk} + S_j\partial_i\beta^j\right\} && 
 \label{SetGRHD}
 \end{eqnarray} 
 where $D=\rho W$ is the relativistic mass density with $W=(1-\text{v}_i\text{v}^i)^{-1/2}$ being the Lorentz factor of the fluid. The relativistic inertia of the fluid is $\xi=\rho hW^2$ 
 while $S_i=\xi\text{v}_i$ is the relativistic momentum  where $\text{v}_i$ is the covariant coordinate velocity of the fluid ($u_i=W \text{v}_i$) normalized to the speed of light $c$.   
 For simplicity and robustness we choose to replace the energy conservation by a simple polytropic law linking the thermal pressure to proper gas density, 
 namely $P=C_o\rho^{\tilde{\gamma}}$ where $C_o$ and $\tilde{\gamma}$ are two positive constants. 
   Relativistic gas thermodynamics prevents the use of a standard polytropic equation of state linking the internal energy of the gas to its density. Following \citet{Meliani04} and \citet{Mignone07}, 
   we can derive such relation by considering the properties of the distribution function of a relativistic gas \citep{Taub48,Mat71}. This leads to  an expression of the internal energy density $U$  as
  \begin{equation}
  U = \frac{P}{\Gamma -1}+\sqrt{\displaystyle\frac{P^2}{(\Gamma -1)^2}+\rho^2c^4} \nonumber
  \label{Eq:internalU}
   \end{equation}   
  where $\Gamma=C_P/C_V$ is the specific heat capacities ratio. Then enthalpy can be directly computed from the previous equation since 
  \begin{equation}
  \rho hc^2 = \frac{1}{2}\left\{(\Gamma+1)U-(\Gamma-1)\frac{\rho^2c^4}{U}\right\}\nonumber
  \label{Eq:Enthalpy}
   \end{equation}  
We developed a GRHD version of the MPI-AMRVAC code that solves the aforementioned conservative equations and thus provides the temporal evolution of conservative variables $(D,S_i)$. 
The structure of the code remains fully parallel by the use of an octree data setting and does conserve the same scaling properties as in \citet{Porth14}. 
   Solving the set of GRHD conservation equations requires to have access simultaneously to both conservative $(D,S_i)$ and primitive $(\rho, \text{v}_i, W, \xi)$ variables. 
   Switching from one set of variables to the other can be achieved by solving a non-linear equation relating the Lorentz factor to the gas enthalpy, namely
   \begin{equation}
   f(W)=D^2 h^2 c^4(W^2-1) - S_iS^i=0\nonumber
   \end{equation} 
The use of a Newton-Raphson algorithm is then possible to get the zeros of such function with a quadratic efficiency. Such procedure must be repeated for every cell of the computational 
domain at ever time step, leading to a slight increase of the computational time compared to classical computations. 

\subsubsection{GRHD code validation}
\label{bondi}

In this paper we focus on  non-spinning black holes as described by the  Schwarzschild metric shown in Eq.~(\ref{Eq:BHmetric}). Such metric is stationary and exhibits a spherical symmetry. The inclusion of general relativistic components into the special relativistic MPI-AMRVAC code then requires to test the ability of the code to handle the radial motion of matter influenced by the gravity of the black hole. One simple but efficient test is to describe the general relativistic Bondi accretion regime \citep{Hawley84}. In such test, one considers the pressureless free-falling motion of matter along a radial geodesic. Neglecting pressure prevents any interaction between the different elements of matter composing the flow. This flow is then  only subject to the gravitational force generated by the central body. Regarding our polytropic equation of state, we achieve such condition by imposing $C_o\ll 1$. 

\begin{figure}
\includegraphics[width=0.48\textwidth]{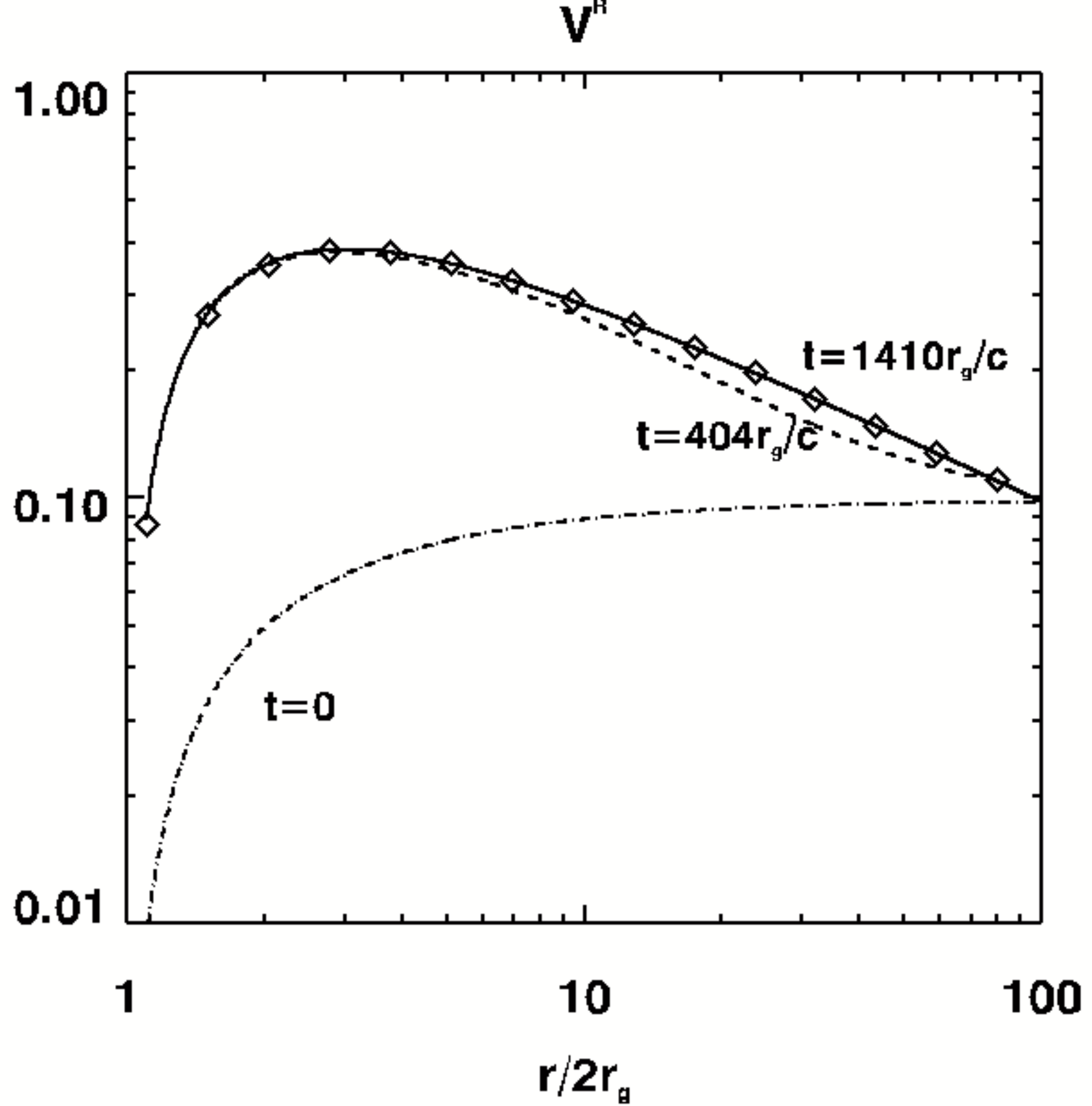}
\includegraphics[width=0.48\textwidth]{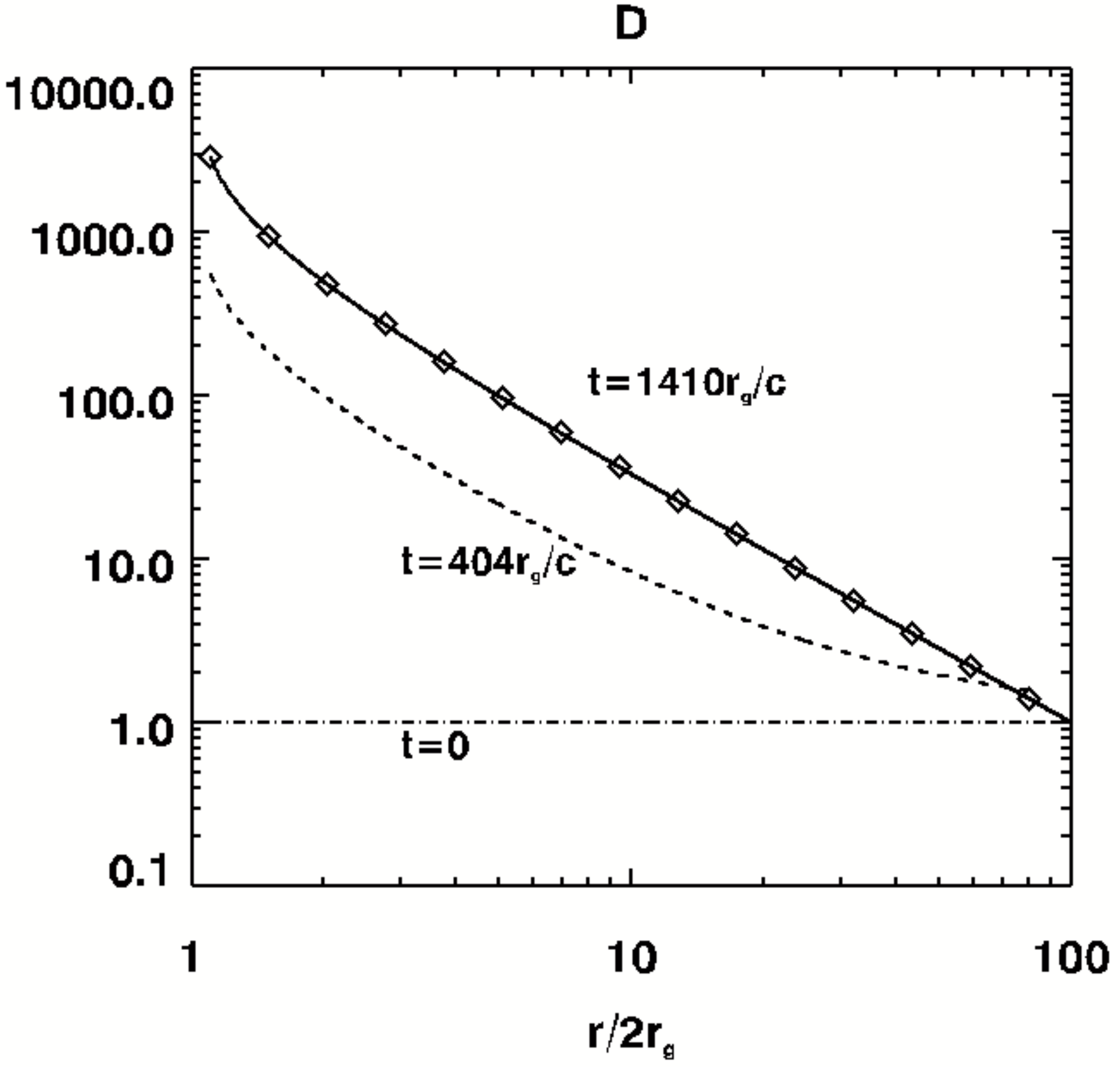}
\caption{Radial contravariant velocity and relativistic density of a spherically symmetric pressureless flow free-falling onto a non-rotating black hole. Both quantities reach a steady-state matching the Bondi accretion regime described in Sect.\FC{\ref{bondi}  and represented by the} diamond symbols.}
\label{fig:Bondi}
\end{figure}

Radial geodesics in Schwarzschild spacetime are described by a very simple equation, namely $du_t/ds=0$. This is the expression of the conservation of the binding energy of the fluid element 
while free-falling onto the black hole. The normalization of the 4-velocity of matter $u_\mu u^\mu=-1$ provides the expression of the contravariant radial velocity $\text{v}^r$ while the mass conservation provides the expression of the relativistic density $D$, namely \citep{Hawley84}
\begin{eqnarray}
\text{v}^r(r)&=&\sqrt{\frac{2r_g}{r}}\left(1-\frac{2r_g}{r}\right) \nonumber \\
D(r) &=& \frac{D_o}{r^2\sqrt{\displaystyle\frac{2r_g}{r}\left(1-\frac{2r_g}{r}\right)}} \label{Eq:Bondi}
\end{eqnarray} 
We set up a spherically symmetric GRHD simulation of such flow where we initially tune the density \FC{and the radial velocity $\sqrt{v_rv^r}$} to  constant \FC{values while the} orthoradial and azimuthal velocities are nul. The computational domain ranges from $2.2r_g$ to $200r_g$ while setting $C_o=10^{-10}$ in order to render the thermal pressure negligible compared to the mass energy. We ran the simulation up to time $t=1410\   r_g/c$ where the simulation reaches a nearly perfect steady-state (residual simulation drops below $10^{-11}$). On Figure (\ref{fig:Bondi}) we display the \FC{radial} evolution of both density and radial velocity \FC{at different times}, showing that they converge toward the Bondi solution \FC{represented by the diamond symbols}. The relative overall variance between the final stage of the simulation and the Bondi solution is smaller than $0.1\%$ which proves that the simulation has converged toward the expected solution. 
\section{Setup and diagnostics for the RWI}
\label{sec:setup}

\subsection{Numerical setup}	

  \begin{figure}
\includegraphics[width=0.48\textwidth]{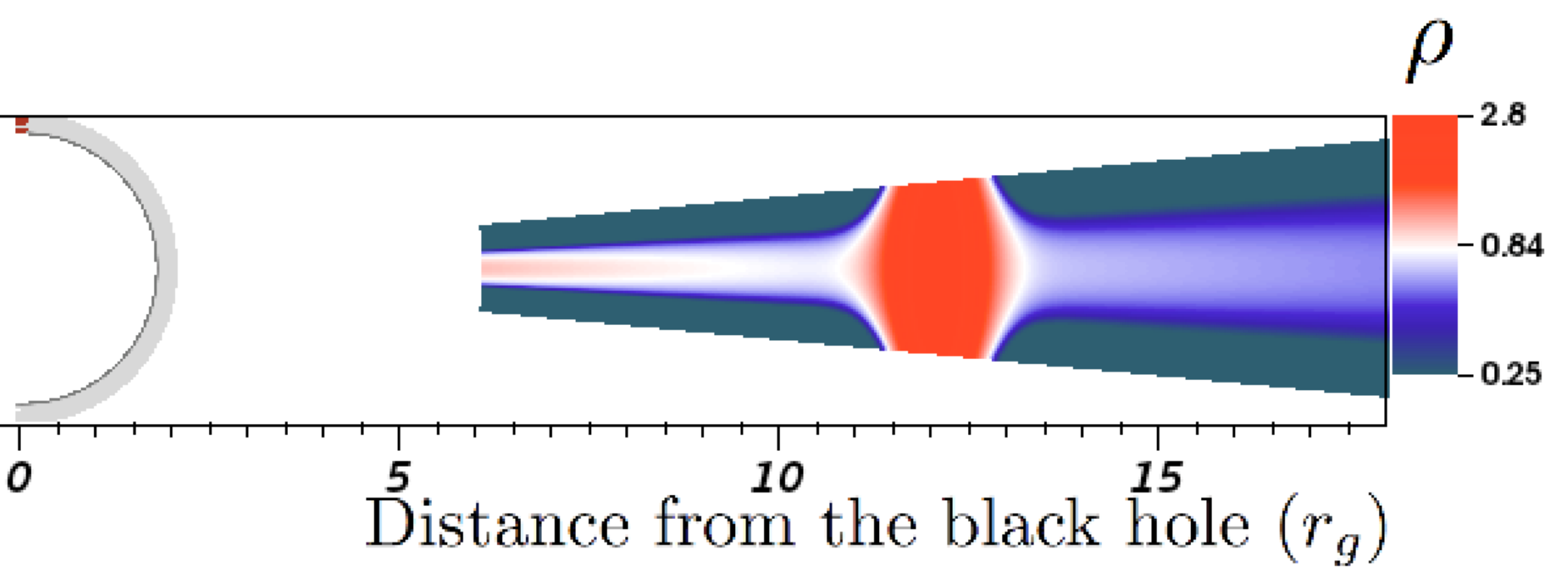} 
 \caption{Initial density distribution in the $(r,\theta)$ plane of an accretion disc harboring a density bump at $r_c=12r_g$. 
  }
 \label{fig:Init_3D}
\end{figure}

In order to trigger and follow the growth of the RWI inside a rotating disc, we have to design initial conditions exhibiting an 
extremum of the vortensity while achieving a perfect balance between the various forces acting on the gas. 
We choose the simplest setup possible for such disc, namely having a localized enhancement of the density or bump\footnote{While mostly used for simplicity and the cleanness of the results, those could have for origin
    some blob of matter falling onto the system and getting circularized before getting accreted as it was proposed for Sgr A$^\star$.}.
    We therefore choose the following radial density profile $\Sigma$ 

\begin{equation}
\Sigma (r)= \Sigma_o\sqrt{\displaystyle\frac{6r_g}{r}}\left(1+3\exp\left\{-\left(\frac{(r-r_c)}{2\Delta}\right)^2\right\}\right)
\label{Eq:Densdistri}
\end{equation} 
where $r_c$ is the value of the corotation radius and $\Delta $   is the typical width of the density bump.  In the context of general relativistic hydrodynamics, the radial equilibrium is enforced by tuning the rotational momentum  $S^\varphi$ 
  such that the radial projection of the centrifugal force balances both the radial gravitational force and the thermal gradient. 
  On the other hand, there is no gravitational force acting on the gas in the $\theta$ direction which requires a balance between the $\theta$-component of the centrifugal force and the pressure gradient component along the same direction. The 3D setup of the disc is then the solution of the following implicit system
  \begin{eqnarray}
\partial_\theta P &=& \alpha S^{\varphi 2} \cos\theta \nonumber \\
S^\varphi(r,\theta) &=&\sqrt{\frac{\xi}{r}\left(\xi\frac{\partial_r\alpha}{\alpha}+\partial_rP\right)}
\end{eqnarray} 
that we solved numerically. The resulting density distribution can be seen in Fig.\ref{fig:Init_3D} \FC{for the case of $r_c=12 r_g$}.  

In order to complete the initial disc setup, we choose to use a polytropic relation linking the thermal pressure to the density. 	
 The polytropic index is set to $3$ while  $C_o =1.8\times 10^{-4}$. The values of these constants  have been selected
such that the ratio of the sound speed to the rotational velocity at the disc midplane remains constant throughout the disc
 ($c_S/\sqrt{\text{v}_\varphi\text{v}^\varphi}\sim 0.04$).  Such choice is consistent with a thin accretion disc structure 
 except near the bump 
 where the local disc scale height is a significant fraction of its spherical radius. Such thin disc displaying a 
 localized \FC{bloated part}  
 meets all the criteria to trigger a very rapid growth of the RWI \FC{hence} preventing any interference from the boundaries to occur within the computational domain.
 
 \noindent We  seed the instability near the bump by adding random velocity perturbations to the disc equilibrium. 
 \begin{equation}
 \delta\text{v}_r=\xi_{\text{ijk}}\text{v}_\varphi\exp\left(-\left(\frac{r-r_c}{\Delta}\right)^2\right)\nonumber
 \end{equation}
 where $\xi_{\text{ijk}}$ is a random variable verifying $<\xi>=0$ and $<\xi^2>=10^{-10}$ when averaged over the whole computational domain.
 
  The computational domain of all the simulations presented in this paper is $r\in [0.5r_c,1.5r_c]$ and $\varphi\in [0,2\pi]$. The spatial resolution for the 2D simulations is $192\times 600$ cells while full 3D simulations  will be performed using the same resolution and extension in the $(r,\varphi )$ plane and  $\theta\in[0.47\pi,\pi/2]$ with $32$ cells in that direction.
 The boundaries of the  computational domain are periodic in the azimuthal direction while the radial ones are designed to absorb the velocity/density perturbations coming from the random perturbations spread  near the density bump. 
 
  In order to time advance the set of equations presented in the previous section, we used a Harten, Lax and van Leer (HLL) solver linked to a Koren slope limiter \citep{Koren93}. A typical 2D simulation requires approximately ten thousands time steps on $20$ processors and lasts less than an hour.
  
 Three dimensional simulations \FC{takes into account} the full disc structure and thus have to deal with the presence of a low density region above and below the disc. We assume the disc to be embedded in such low density region that should not interfere with the RWI development. In order to design such buffer zone, we impose that below a typical floor density (typically a hundredth of the disc midplane density), no motion is allowed. Typical 3D simulations exhibiting the aforementioned resolution ran for about thirty hours on $20$ processors in order to reach the same physical time than the 2D simulations. 
\subsection{Methodology}
 \begin{figure}
\includegraphics[width=0.5\textwidth]{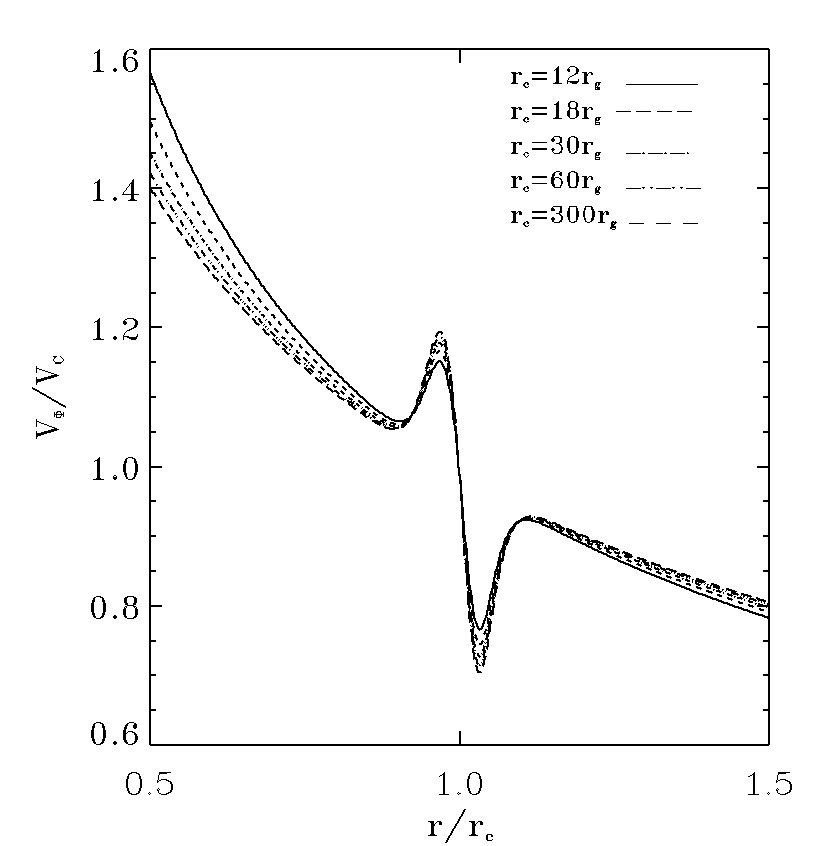}
\caption{Rotational velocity $\sqrt{\text{v}_\varphi\text{v}^\varphi}$ of the disc normalized to its value at the corotation radius for the various 2D initial setup. 
}
\label{fig:vphi_150rs}
\end{figure}

	In this paper, we intent to study the influence of a Schwarzschild black hole on the evolution of the RWI  depending on how 
	close to the compact object this instability  develops in the surrounding disc.  To this end we will perform a set of simulations
	for different corotation radii, namely $r_c \in \{300, 60, 30, 18, 12\}\ r_g$. 
	 In order to compare the different simulations we scaled distances in each simulation with respect to their  
	 corotation radius. Accordingly
	 time will be scaled to $T_c(r_c)$ the rotation period at the corotation radius as measured by a {\it local} observer located at a distance 
	 $r_c$ from the black hole.  These various simulations represent distinct test cases where we study the growth of the RWI occurring at locations closer and closer to the black hole.

\begin{figure}
\includegraphics[width=0.5\textwidth]{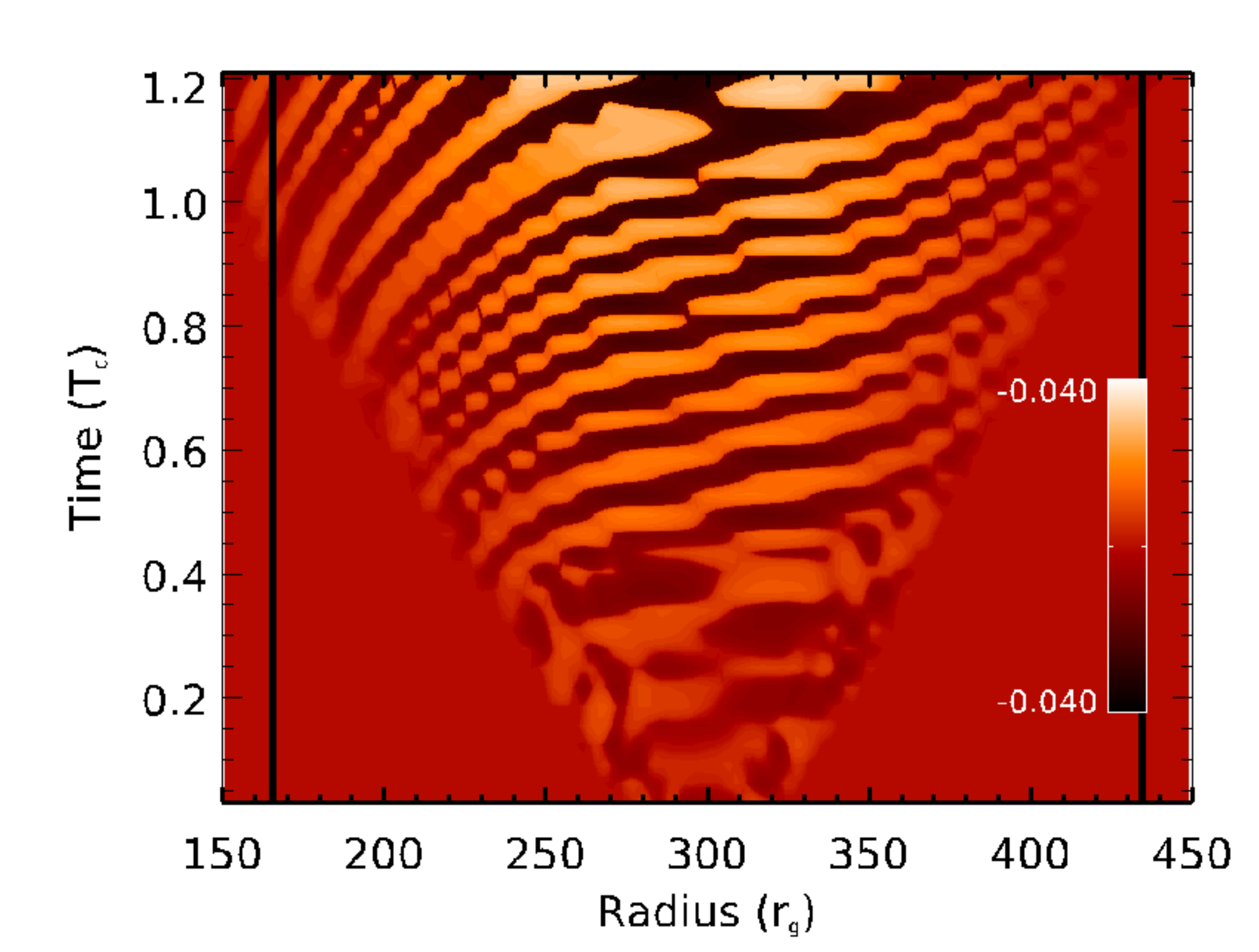}
 \caption{Modified logarithmic colormap of the density perturbations as a function of the radius and time for $\varphi_0=\pi$ at $r_c=300 r_g$ 
 Thick black lines represent the limit of the box on which we did the Fourier transform. Time is normalized with respect to the rotation period $T(300r_g)$ as measured by an observer located at $r_c=300r_g$. }
 \label{fig:RT_150}
\end{figure}

   Fig.\ref{fig:vphi_150rs} shows the behaviour of the rotational velocity scaled to its value at $r_c$ for each simulations. 
       The corresponding velocity profiles are not exactly identical \FC{as the gravity is not an exact power-law.
       However they do not significantly differ one from another. Such similar setup will then allow us to compare the evolution of the RWI in the various simulations. \\
     In order to compare our computations with  previous works, we start our analysis with the simulation 
     having the corotation radius located farthest away from the last stable orbit of the black hole, namely $r_c=300 r_g$.
	For several reasons we have selected a disc configuration where we expect a strong and fast growth of the instability.} First of all it ensures than any parasitic instability 
	will not have time to develop before the RWI reaches its non-linear stage and therefore warrant that we are studying only the RWI. Secondly, it prevents any interaction with 
	the edges of the simulation box 
	as the spiral waves do not reach them before the end of the simulation as can be seen on Fig.\ref{fig:RT_150}.
      This is not an issue when running a simulation with $r_c=300 r_g$ because there is no hard limit for the simulation box. 
	Nevertheless, as we get closer to the last stable orbit of the black hole, we need to ensure
	that we are still confronted to the same boundary conditions. Hence we set the  radial extend of the smallest simulation to $[6r_g,18r_g]$ 
	so that it will  match the relative radial size of the farthest simulation while insuring that the simulation is not influenced by the unstable part of 
	the flow located inside the last stable orbit of the disc.
	

 \subsection{Set of RWI diagnostics: near-Newtonian case  at $r_c=300r_g$}

\begin{figure}
\includegraphics[width=0.5\textwidth]{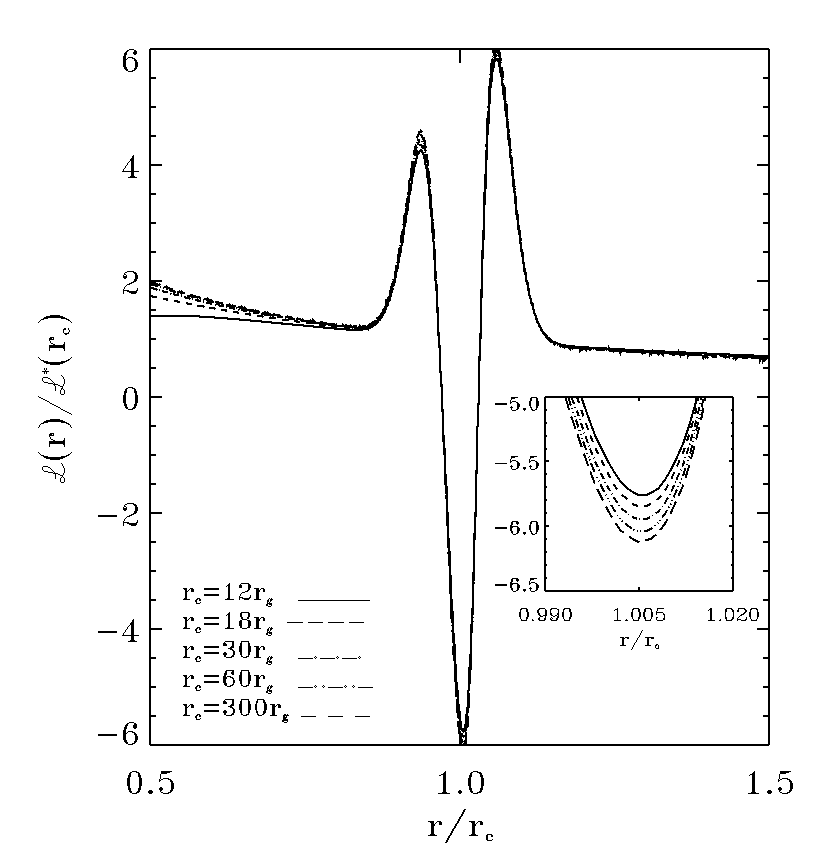}
\caption{RWI criteria ${\cal L}$ for the entire set of $2$D simulations normalized to ${\cal L}^*(r_c)$ where ${\cal L}^*$ corresponds to  the criteria of a disc without any density bump.
The subplot represents a zoom of the minima where we can actually see the small differences in our setups.}
\label{fig:critere}
\end{figure}

      The first simulation considered in this paper, at $r_c = 300 r_g$,  is consistent with a near-Newtonian case. Indeed, that far away from the last stable orbit of the black hole, the gravitational field generated by  the compact object is very similar to a Newtonian one. We can then compare the different features  of this first computation with previous simulations of the RWI performed in a pure Newtonian case.
      In order to be certain that we are actually dealing with the RWI the first thing to check is the criteria in Eq.\ref{eq:critere}.  
      The aforementioned disc setup has been designed solely for that purpose so the criteria is actually fulfilled as can be seen on Fig.\ref{fig:critere}. 
         On top of that we see that the behaviour of ${\cal L}$ is relatively similar for all the simulations in the set with only a difference of 
         about $10$\% near the extremum.
         This will allow us to compare the results and growth rate of the RWI between the different runs.
         \PV{Fig.\ref{fig:critere} also shows that vortensity profiles get flatter as the inner edge of the simulation is getting closer and closer to the last stable orbit around
          the black hole.
         Such flattening denotes the impact of a full general relativistic description of the black hole's gravity upon the velocity.}\\
	Once we know the criterion is fulfilled we can seed some random perturbations around the corotation  and follow how they evolve with time.
	This is what we see on Fig.\ref{fig:RT_150} which shows, stacked in time, the relative density perturbations: 
\begin{eqnarray}
  &&	\frac{\rho(r,\varphi_0,t) -\rho_o}{<\rho(r,\varphi,t)>_\varphi}.
\end{eqnarray}	
	Such plot makes it easy to track propagating waves as time evolve. Indeed, we can  follow  the evolution of the density perturbations and the spiral waves that develop from the corotation zone where seed perturbations were added. This ensures that the evolution is as \FC{expected} for the RWI and
	also that no problem occurs from the interaction with the boundaries of the simulation box. \\
\newline
  \begin{figure}
\begin{tabular}{ll}
\includegraphics[width=0.24\textwidth]{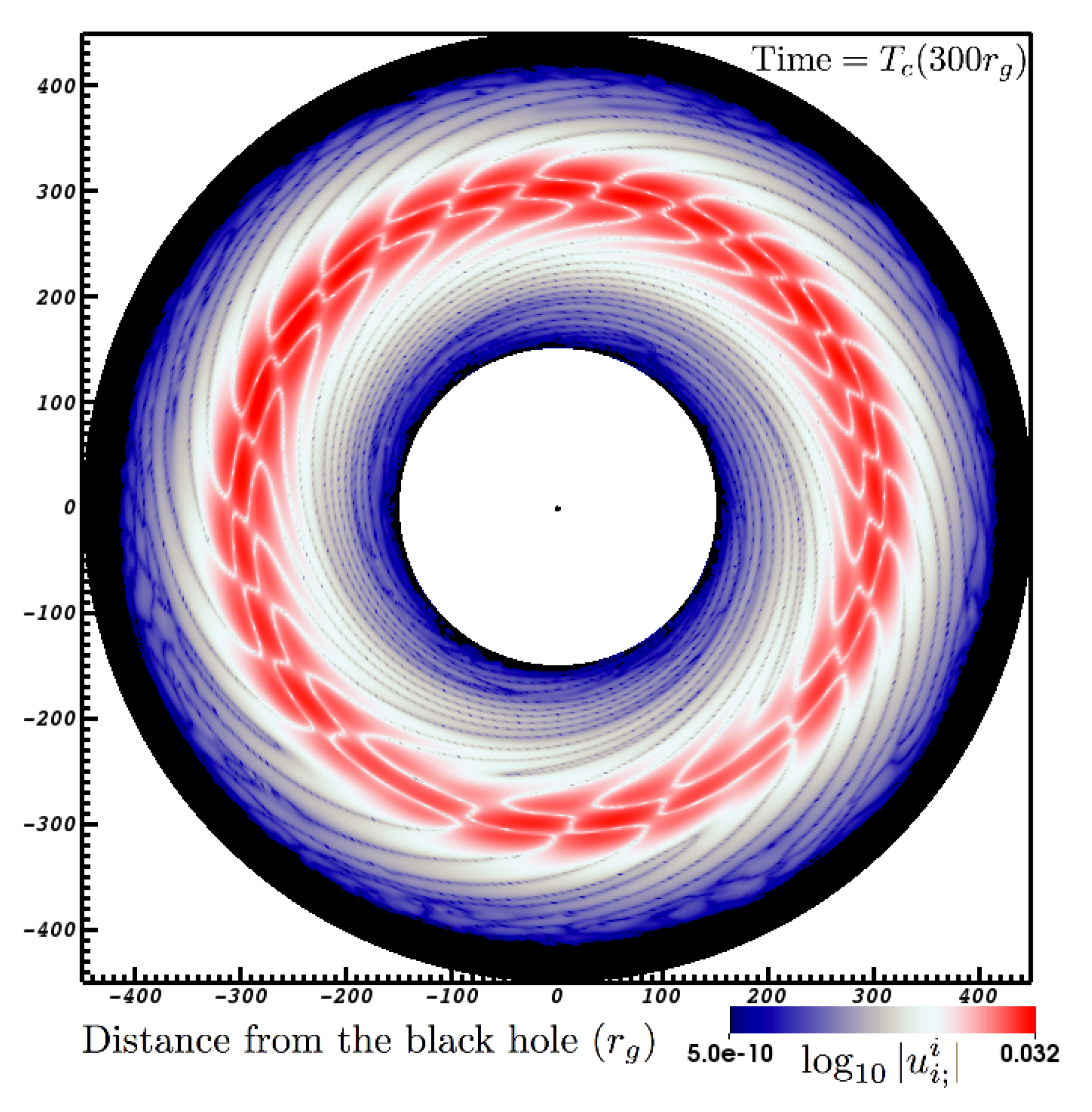}  &
 \includegraphics[width=0.24\textwidth]{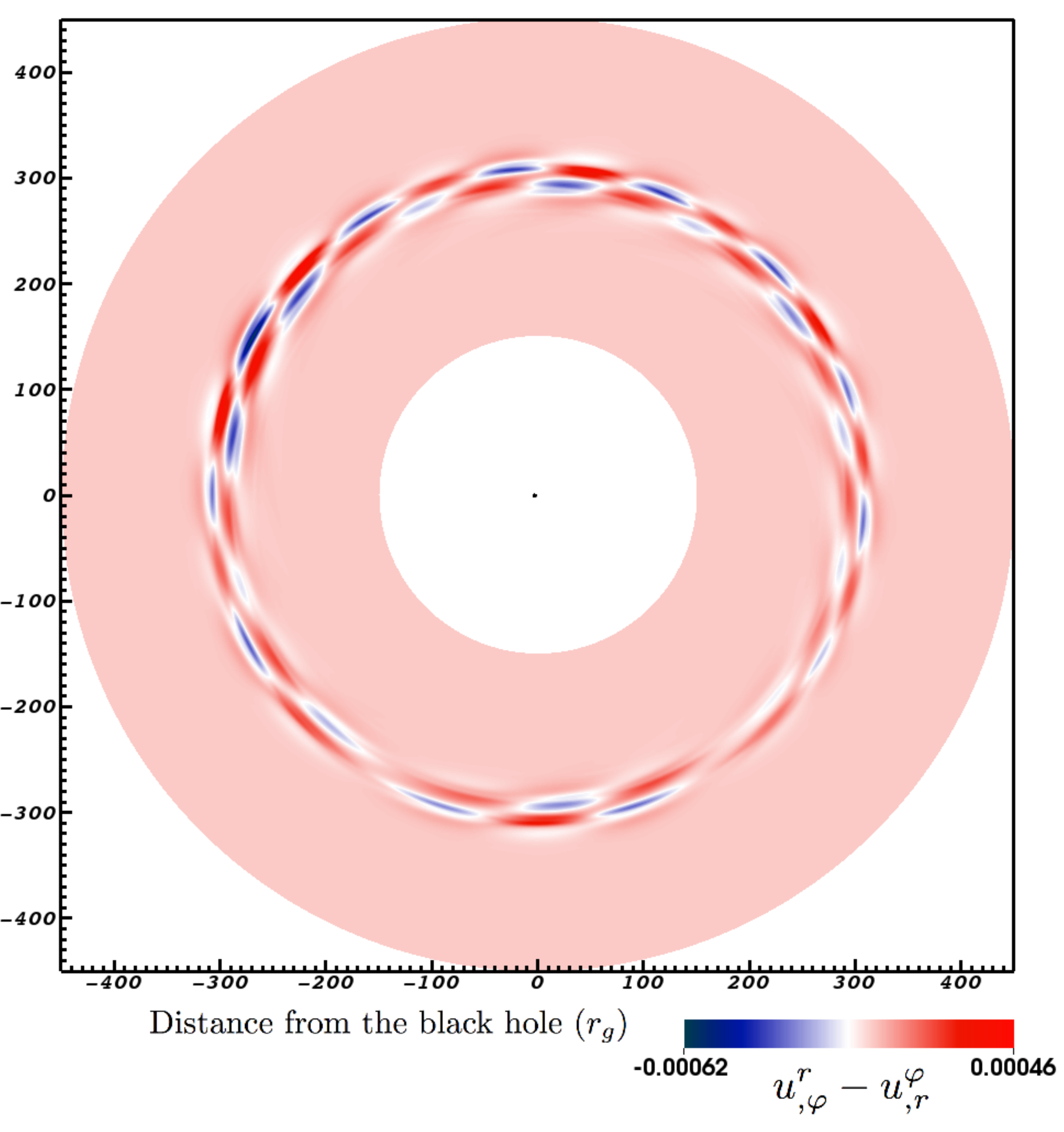} 
 \end{tabular}
 \caption{Logarithmic colormap of the divergence ($u^i_{;i}$) and linear colormap of the  curl ($u^r_{;\varphi}-u^{\varphi}_{;r}$ ) 
 of the velocity field showing the spiral waves and vortices after one orbit at the corotation radius $r_c=300 r_g$} 
 \label{fig:div_1502D}
\end{figure}
    Another way to ensure the identity of the instability is to decompose it into its compressional ($u^i_{;i}$) 
    and torsional ($u^r_{;\varphi}-u^{\varphi}_{;r}$)  components that can be seen on Fig.\ref{fig:div_1502D}. 
    On those plots we can track the evolution of the spiral waves as with the density, though much earlier. The torsional component is also the 
    best way to look at the formation and evolution of the vortices. 
     On those plots the size of the inner disc radius can be inferred  from the size of the central dark spot which represents the event 
     horizon of the the black hole.
    
	We see that for this near-Newtonian case ($r_c=300 r_g$) the compressional and torsional components behave as expected with the spiral waves propagating from the corotation where vortices are also formed. 
 	From the number of vortices we can find out which instability modes are present. While this is easy when there is only a very limited number of 
	modes or when one of them is strongly dominant,
	it generally requires to go to the Fourier space to fully dissociate the various instability modes. From Fig.\ref{fig:mode_compall} we 
	can see that the instability modes strength 
	distribution is not uniform, hinting that there are one or more 
	low $m$ modes getting stronger than the high $m$ modes we get from the start (see \citet{TV06} for a more detailed mode study). 

\section{Impact of a Schwarzschild black hole gravity upon the RWI}
\label{sec:towardLSO} 
 
         Now that we have recovered the near-Newtonian case at $r_c=300 r_g$ we can perform a series of simulations where a similar 
         boated disc is considered with a corotation radius $r_c$  closer and closer to the black hole. We
         choose to run the set with $ r_c \in \{300, 60, 30, 18, 12\}r_g$. For space reason, and the fact that they are very similar, we present only 
         the diagnostics for the case $r_c=12 r_g$.
         
\subsection{Comparison in real space}    

\begin{figure}
\begin{tabular}{cc}
\includegraphics[width=0.25\textwidth]{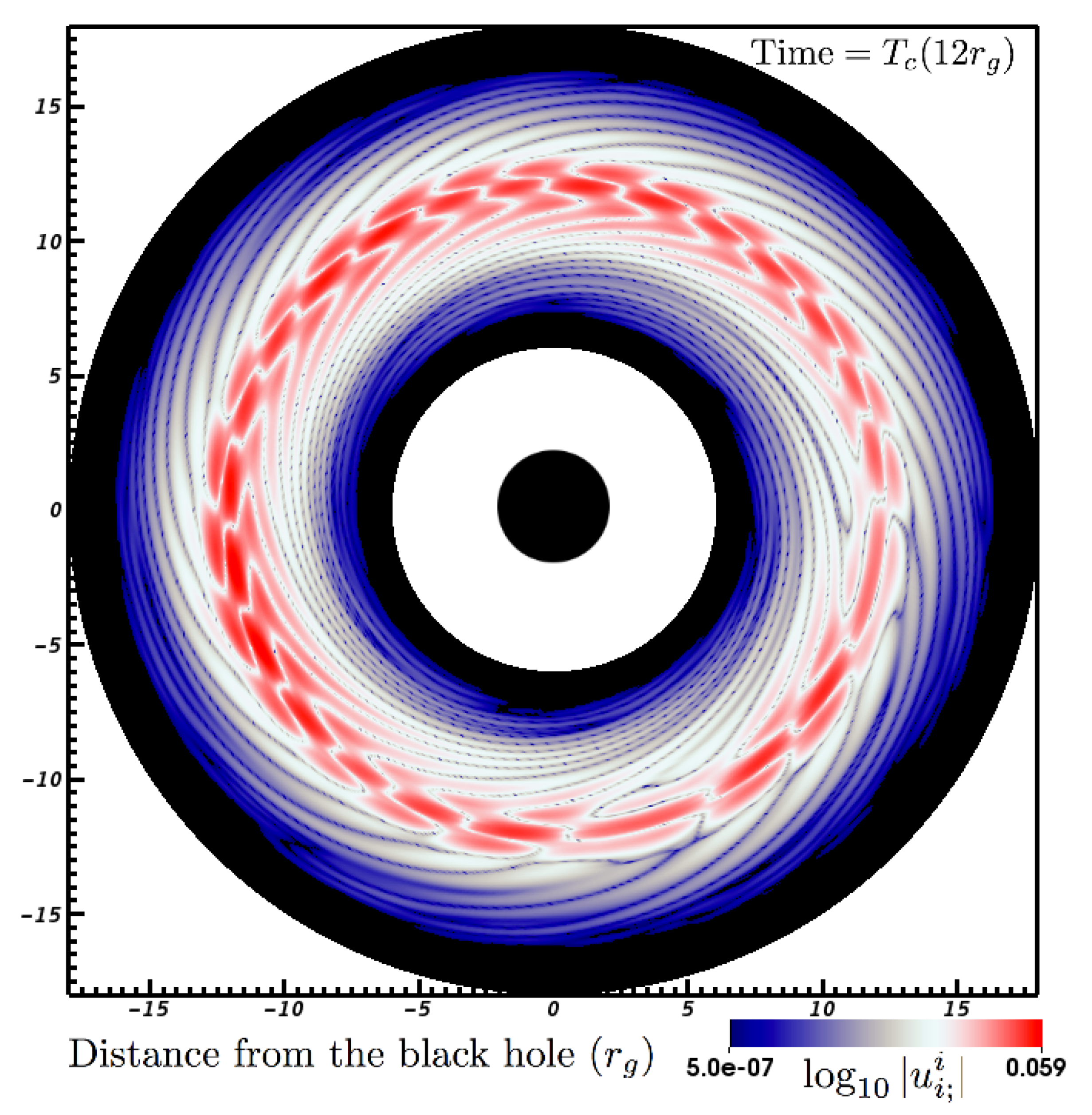}  &
 \includegraphics[width=0.238\textwidth]{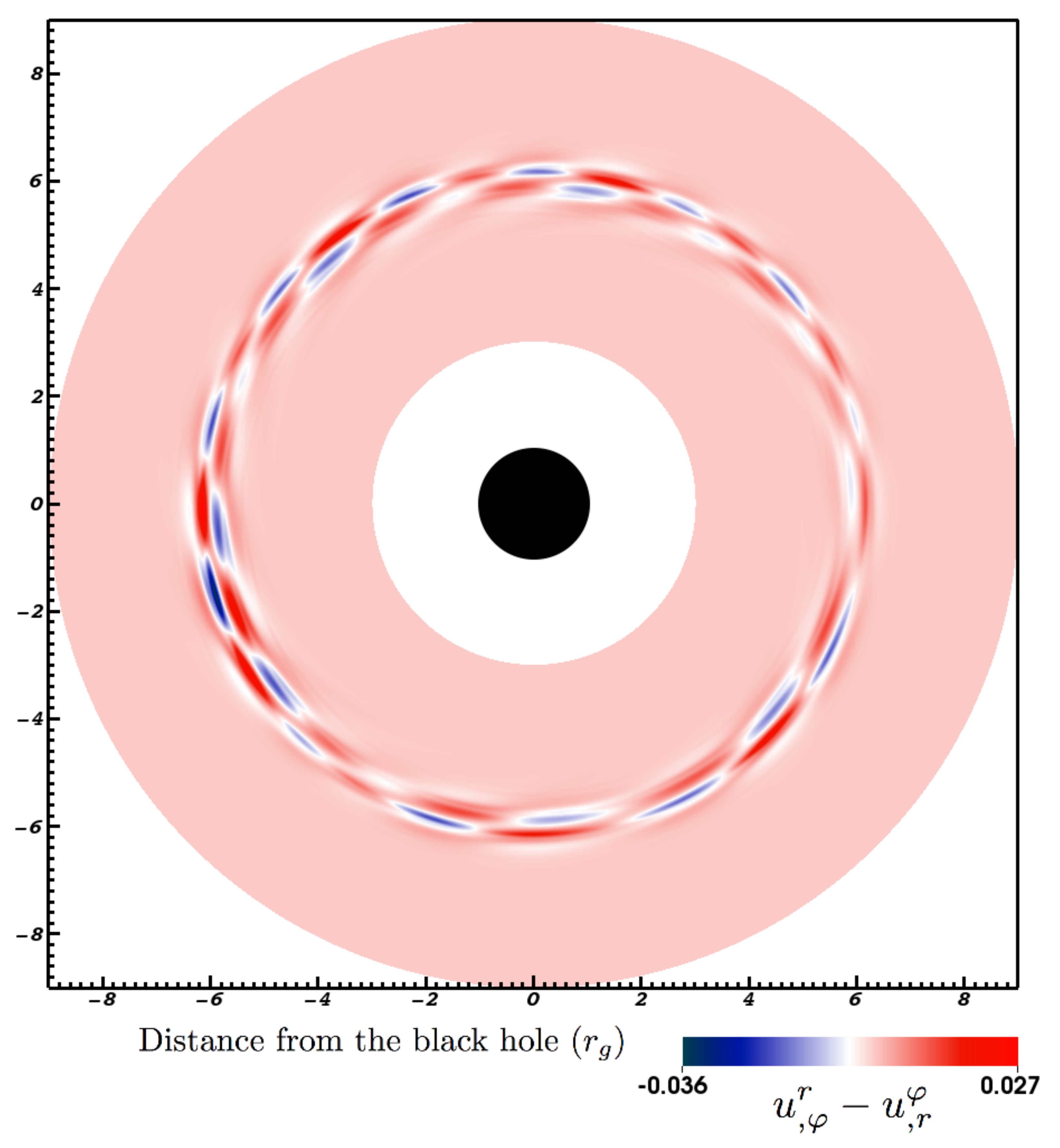} 
 \end{tabular}
 \caption{Logarithmic colormap of the divergence ($u^i_{;i}$) and linear colormap of the  curl ($u^r_{;\varphi}-u^{\varphi}_{;r}$ ) 
 of the velocity field showing the spiral waves and vortices after one orbit at the corotation radius $r_c=12 r_g$} 
 \label{fig:div_2D}
\end{figure}
	When looking at the simulation in the real space, namely the $(r,\varphi)$-domain, we obtain very similar diagnostics as we did for the near-Newtonian case. For example,
	we see on Fig.\ref{fig:div_2D} the compressional and rotational components of the velocity  for the simulation at  $r_c=12 r_g$ which is to be compared with  Fig.\ref{fig:div_1502D}
	where the corotation was twenty five times further away from the black hole (the \lq black circle\rq\ at the center is hence 25 times smaller).
	We see that the main behaviour of the RWI is unaffected by its closeness to the black-hole, its overall behaviour being indistinguishable from the near-Newtonian case. 

\begin{figure}
\includegraphics[width=0.48\textwidth]{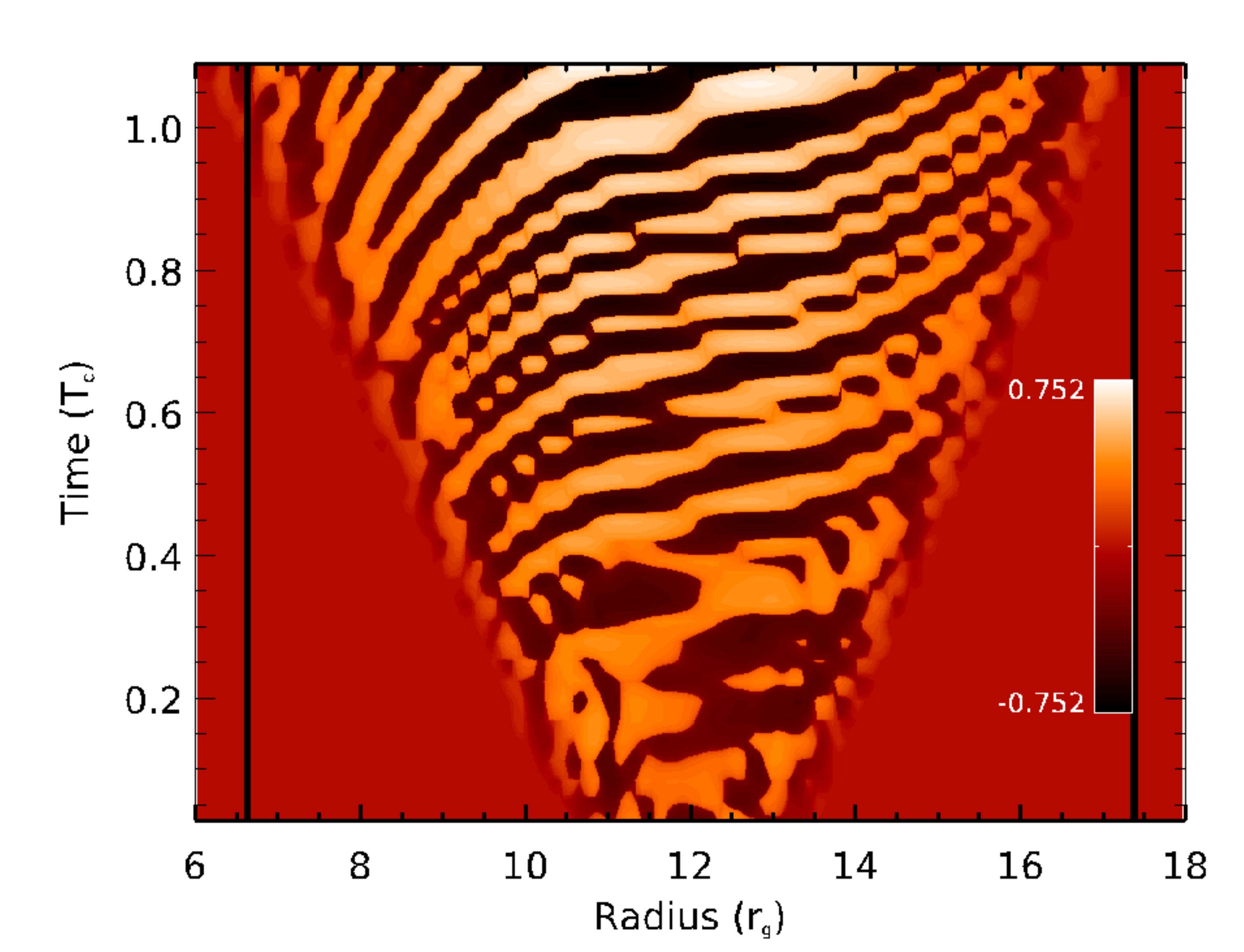}
 \caption{Modified logarithmic colormap of the density perturbations as a function of the radius and time for $\varphi_0=\pi$ at $r_c=12 r_g$ 
 Thick black lines represent the limit of the box on which we did the Fourier transform. Time is normalized to the rotation period $T(12r_g)$ as measured by an observer located at $r_c=12r_g$.}
 \label{fig:RT_6}
\end{figure}     
	Nevertheless, we detect some slight differences when looking more closely at the density perturbations and especially their time evolution as seen in Fig.\ref{fig:RT_6}.  Indeed, according to our simulation design, the inwards sound wave (generated by the initial perturbations) should reach the inner border after one local orbital time $T_c(r_c)$ while the outwardly propagating sound wave will reach the outer border after approximately $1.27\ T_c(r_c)$. Our simulation setup makes these values independent of $r_c$. Looking at Fig.\ref{fig:RT_150}, we recover this feature but not on Fig.\ref{fig:RT_6} where both waves seems to reach the borders approximately at the same time. This discrepancy stems from a general relativistic local time dilatation occurring as one gets closer and closer to the black hole.  Indeed general relativity shows that the clock rate $d\tau/dt$ of a referential located at distance $r$ from a Schwarzschild black hole is provided by the lapse function, namely $d\tau/dt=\alpha(r)$.
In contrast to the near-Newtonian simulation where the lapse function $\alpha$ is very close to unity over the whole computational domain, the simulation having $r_c=12r_g$ exhibits a radially increasing $\alpha$  that leads to a stretching (along the time axis) of the left part of the plot as time dilatation is more significant on the inner side of the corotation radius than on its outer side.
 One has to keep in mind at this stage that the velocity and density maps presented in the previous figures are {\it not} the image that a Newtonian observer would perceive since light emitted from the disc is also subject to a relativistic distortion induced by the black hole gravity. 
 
\subsection{Comparison in the Fourier domain}
	  \begin{figure}
\begin{tabular}{c}
\includegraphics[width=0.48\textwidth]{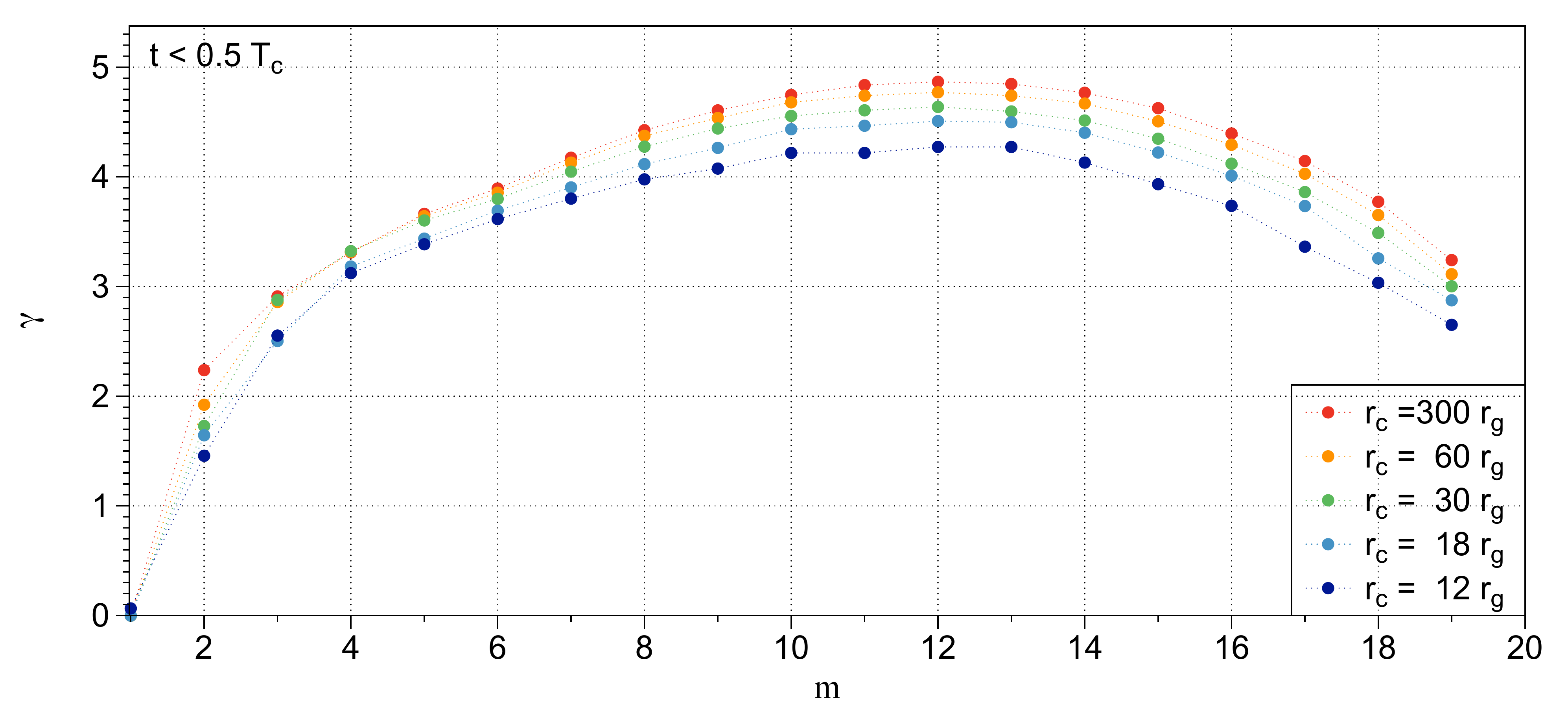} \\
 \includegraphics[width=0.48\textwidth]{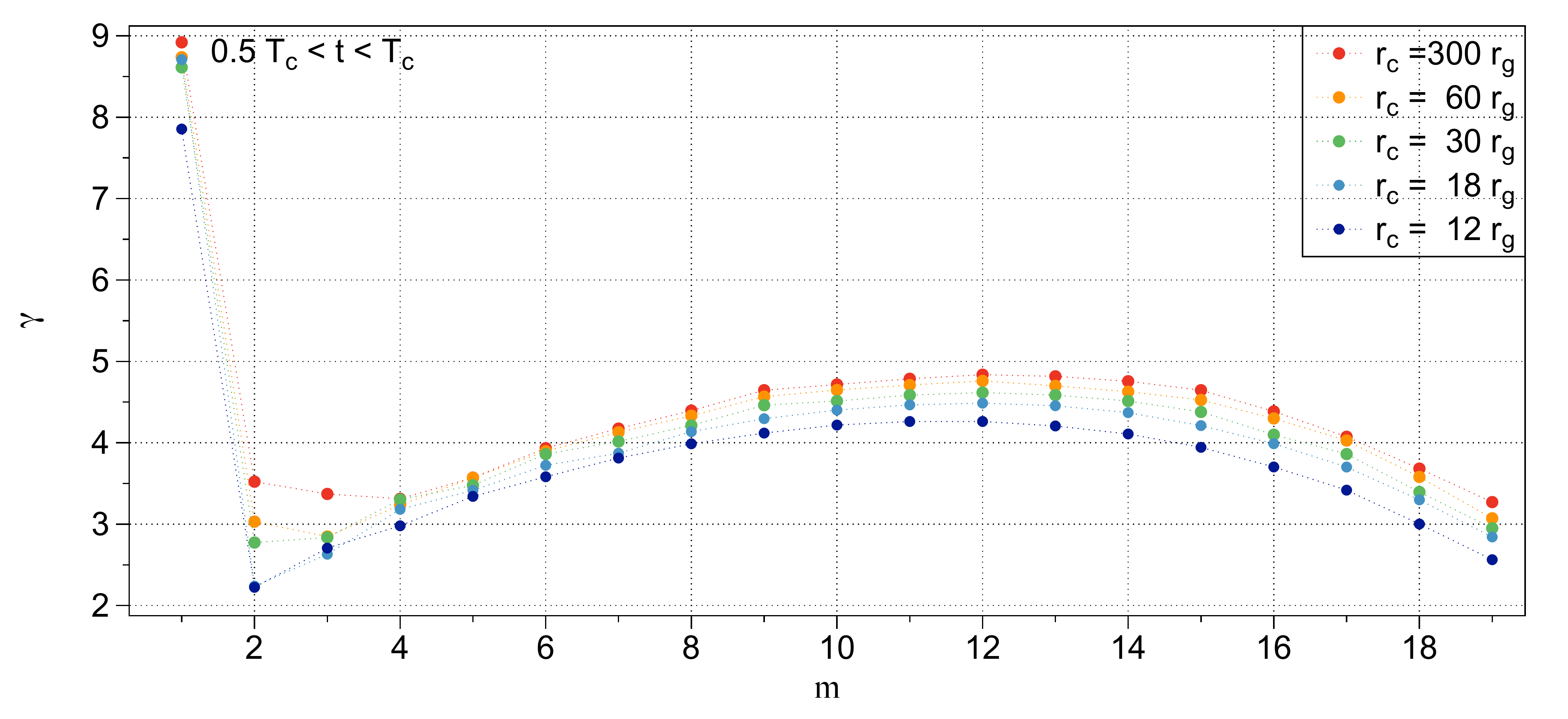}
\end{tabular}
 \caption{Growth rate $\gamma$ of the first nineteenth azimuthal modes for every location of the corotation radius $r_c$ studied in $2$D. 
 The top panel stands for growth rates measured during the initial linear phase of the instability while the bottom one displays growth rates measured during the second half of the simulation.  } 
 \label{fig:mode_compall}
\end{figure}
	Since all the plots presenting physical quantities in the real space domain are so similar, we have decided to study the evolution of the growth rate 
	of the first few instability modes in the Fourier domain. 
	Indeed, it is an interesting observable diagnostic as the presence of a perturbation mode in the disc would be translated as a flux modulation 
	which could be observable\footnote{Indeed, the fact that several low-modes can have a high enough growth rate to be detectable 
	has been proposed as an explanation to apparent multiplicity of 
	frequency related HFQPO in microquasars \citep{TV06}.} Here we will be looking at the growth rate of those modes to see if being closer to the last stable orbits 
	influences what we would observed. 	
	
	For each simulation of our set we have computed the Fourier transform of the rest density $\rho$
	inside the box that encompass the entire span of the spiral wave up to one orbit at the corotation. The boundary of the box is visible on Fig.\ref{fig:RT_150}. and Fig.\ref{fig:RT_6}. and is similar
	for all the simulation in between.
\begin{figure*}
\begin{tabular}{c}
\includegraphics[width=0.95\textwidth]{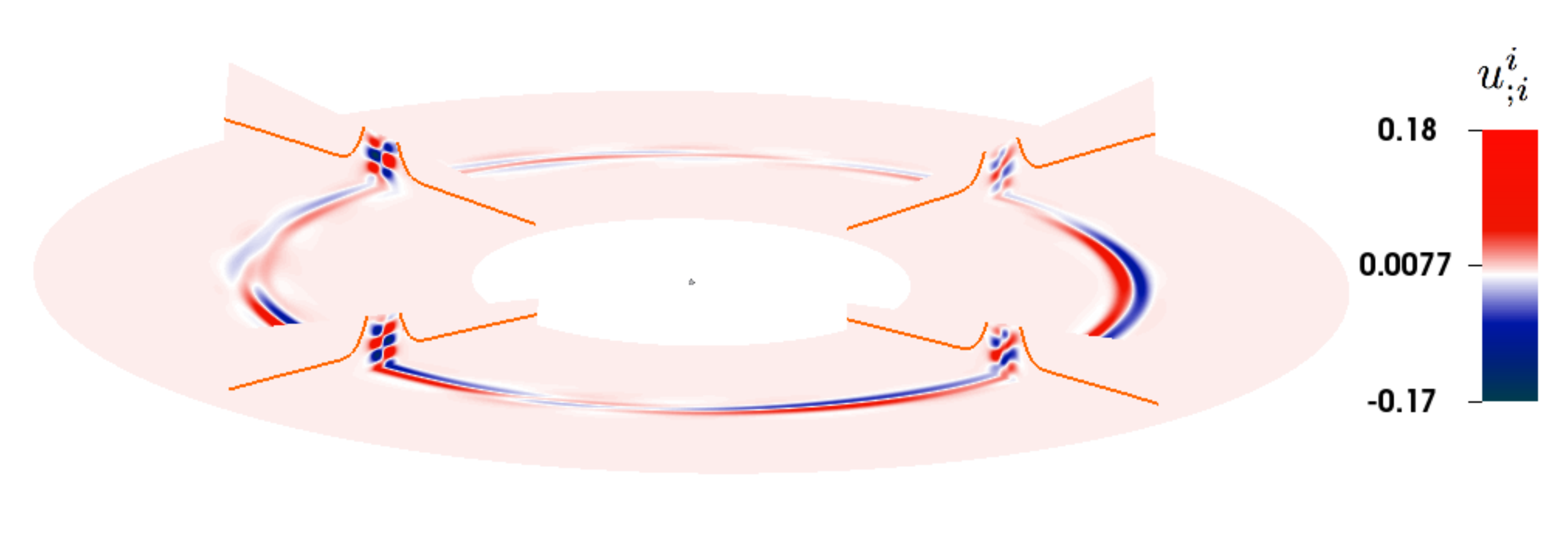} \\
\includegraphics[width=0.95\textwidth]{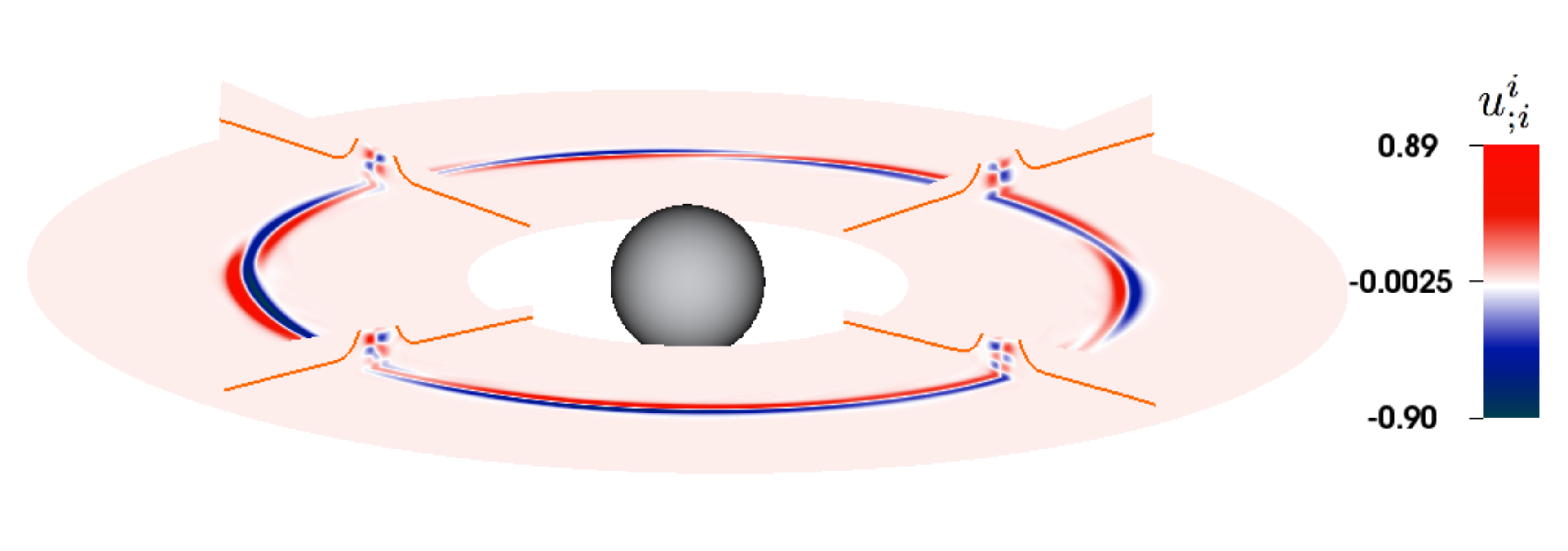}
\end{tabular}
 \caption{
 Linear colormap of the divergence ($u^i_{;i}$)  at  $r_c=300r_g$  and $r_c=12r_g$. } 
 \label{fig:Div_3D}
\end{figure*}
\begin{figure*}
\begin{tabular}{cc}
\includegraphics[width=0.48\textwidth]{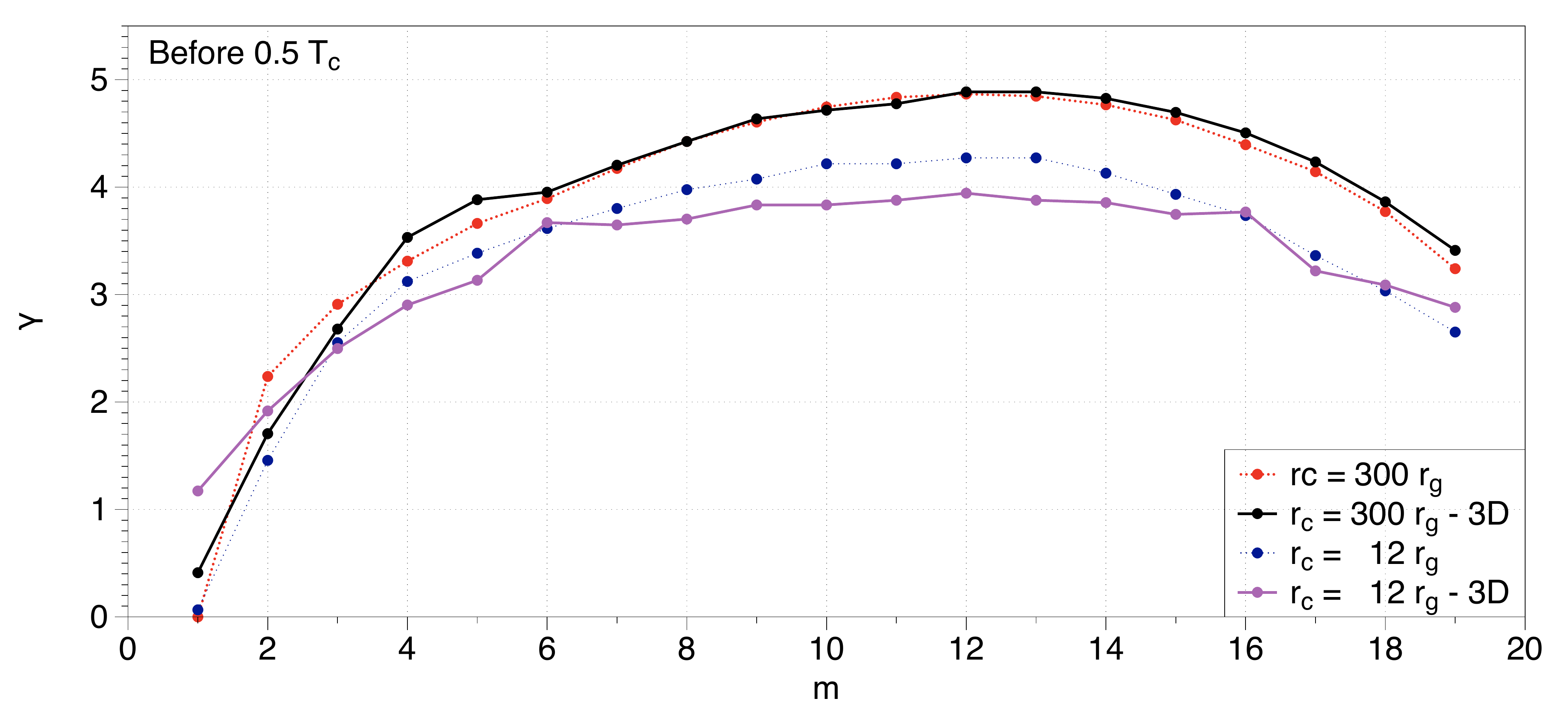} & \includegraphics[width=0.48\textwidth]{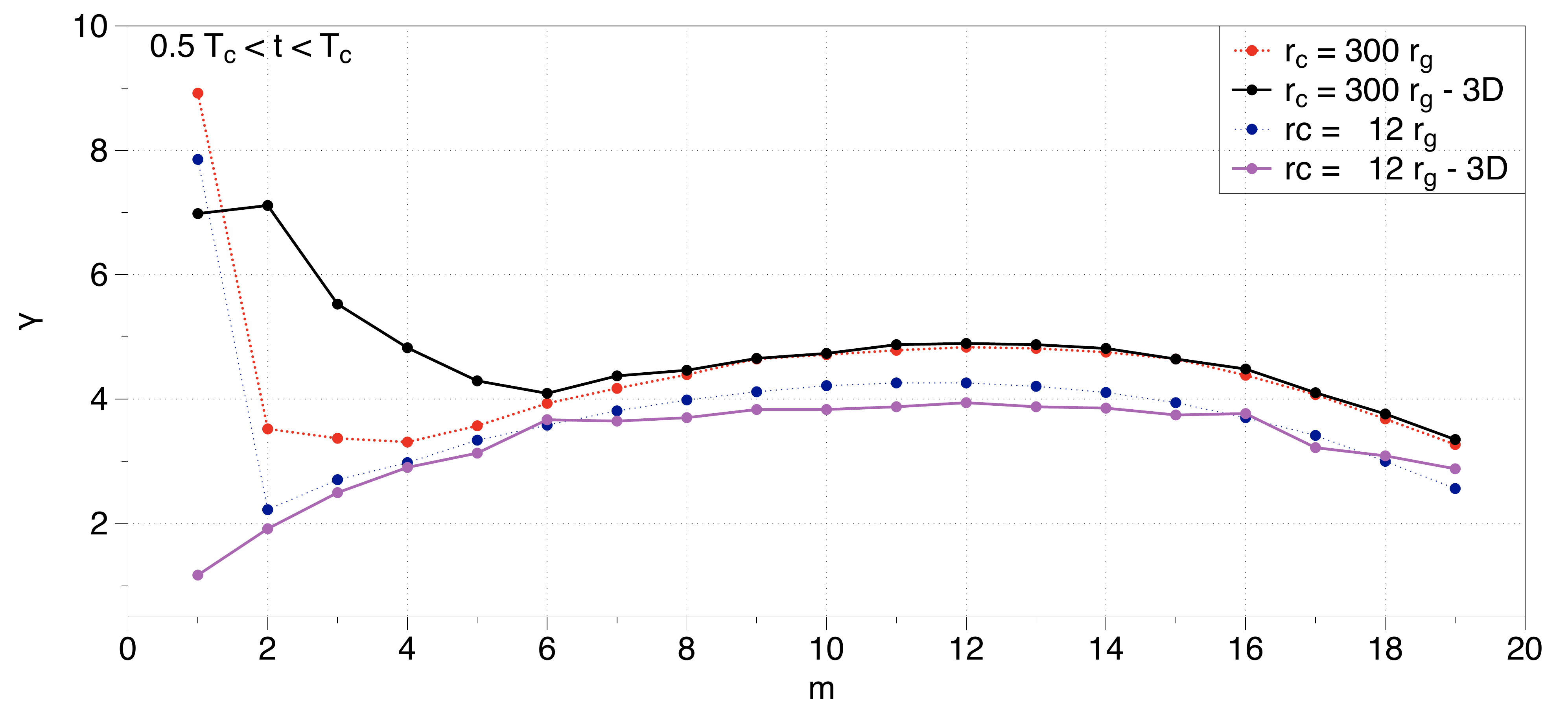}
\end{tabular}
 \caption{Growth rate $\gamma$ of the first nineteenth azimuthal modes for $r_c=12r_g$ and $r_c=300r_g$ computed from both the $2$D and $3$D simulation.
 The left panel stands for growth rates measured during the initial linear phase of the instability while the right one displays growth rates measured during the second half of the simulation. } 
 \label{fig:mode_compall3D}
\end{figure*}	

	On Fig.\ref{fig:mode_compall}.  we choose to display the measured growth rate of the first nineteenth azimuthal modes of the instability for all the simulations of our set, meaning  
	$ r_c \in \{300, 60, 30, 18, 12\}r_g$.  On the top we have the early behaviour of the RWI mode, namely before $T_c/2$. This confirms what we can see directly on the plot of the divergence
	and curl of the velocity Fig.\ref{fig:div_1502D} and Fig.\ref{fig:div_2D}, namely that high $m$ modes are dominant in the disc with the maximum growth rate in this early
	rise of the instability being for the mode $m=12$ closely followed by the mode in between $10$ and $14$. 
	
	As time evolves we see on the bottom panel of  Fig.\ref{fig:mode_compall}. that, while the high $m$ modes  stay at the same level as they were originally, the lower $m$ modes
	becomes dominant. Ultimately this will lead to the merging of the vortices,  as it has been observed in all the simulations of the RWI that were performed for a longer period
	\cite[see for example][]{VarnT06,Vin13}. Here we cannot push the simulation that far as the bump was put {\em by hand} and will be eventually destroyed by the instability.
	The study in the Fourier space allows us to recover the start of this merging with the low-$m$ mode having the strongest growth rate as early as one 
	orbital period. 
	The most interesting results of Fig.\ref{fig:mode_compall} is the overall similarity of the behaviour as the corotation gets closer to the last stable orbit of the black hole. 
	While this shows that, even in a full GR simulation, the basic features of the RWI stay the same, we do see a decrease in the growth rate. 
	While we try to keep the disc structure, Fig.\ref{fig:vphi_150rs}, and instability criteria, Fig.\ref{fig:critere}, as similar as possible between all the simulations, there is a net decrease
	as the corotation radius gets closer to the black hole. It is difficult to untangle this change in the initial conditions from a potential GR effect.
	
\section{RWI in fully 3D accretion discs near a non-rotating black hole}
\label{Sec:3Dsimus}
	It was shown previously that the $2$D and $3$D behaviour of the RWI was similar enough to warrant the use of 2D simulation when comparing with observation until
	we get better observational data \citep{Vin13}. But this work was done in the pseudo Newtonian potential approach. Here we have the opportunity to test this results without approximation.	\\
	The setup we took is very similar to the one we took for the 2D case as we aim to compare the results of both the 2D and 3D simulations. This setup is presented in Section \ref{sec:setup}
	and Fig.\ref{fig:Init_3D} \FC{depicts the mass density of the disc} with $r_c=12 r_g$. As the 3D simulations are both more time and space consuming, we choose to focus only on the near-Newtonian case
	at $r_c=300r_g$ and the case where the inner edge of the disc is at the last stable orbit $r_c=12 r_g$. Especially after showing how similar are the 2D runs, both extreme cases will give us a good view of the 3D behaviour.\\
	Again, we start comparing the simulations in the real spatial domain, looking, for example in Fig.\ref{fig:Div_3D}, at the behaviour of the divergence for both the case at $r_c=300 r_g$ and
	$r_c=12 r_g$.  Note that this time the colormap is linear in order to identify the contours of the various vortices (at the disc mid plane and along the bump height)  and therefore it cannot be compared directly to  Fig.\ref{fig:div_1502D} and Fig.\ref{fig:div_2D}. 
	We see once again that the behaviours of the two extreme simulations of our set are very similar. \\	
	We also identify the behaviour already noted in \citet{Vin13} corresponding to the 3D simulations having dominant low $m$ modes earlier than it is in 
	the 2D simulations.
	Indeed, we did recover this behaviour of merging vortices toward low $m$ modes in a few of the 2D simulations we ran longer than one orbit but choose to not exploit the data here as
	the bump was starting to be modified by the instability and the results could not be easily compared. \\		
	In order to look at this behaviour in more details we performed the same time-domain analysis as in the 2D case, this time doing the Fourier transform between $\theta \in [0.49\pi , 0.5\pi]$
	which represent most of the height of the disk without the bump. We choose to cut out part of the bump in our analysis as having the entire bump would also add a large part of \FC{the low density region}.
	Not having the entire bump means that some modes are underestimated but this way we do not have any interferences from the \FC{low density region} in the Fourier transform.
	Fig.\ref{fig:mode_compall3D} shows the comparison between the growth rate $\gamma$ of the first nineteenth azimuthal modes for $r_c=12r_g$ and $r_c=300r_g$ computed 
	from both the $2$D and $3$D simulations. We see that the overall behaviours between the 2D and 3D runs are in good agreement. 
	In the early phase of the instability
	the growth rates of the 3D simulation computed in the disc only roughly matches the 2D ones. As previously said the slightly lower growth rate for a few 
	modes is understandable as we have artificially cut part of the
	bump out of our Fourier transform, so the mode living there do not have their full growth rate.
	For the growth rate computed between $0.5 T_c$ and $T_c$, we see, as was already noted by  \citet{Vin13},  that the low-$m$ modes 
	in the 3D run are stronger than in the 2D cases. 
	Indeed, those 3D modes are much closer to the non-linear evolution we see in 2D shortly after one orbit when the vortices 
	have had a little more time to merge. 
	Despite the fact that 3D simulations are closer to have the full development of the instability, the resulting differences between 2D and 3D computations 
	remain small. 2D simulations are then likely to stand as a good approximation of the RWI evolution, especially in the context of comparing with 
	observations from available instruments whose spatial resolution prevent us from distinguishing small scale details in the disc structure.
\section{Conclusions and observational consequences}
\label{Sec:conc}
      The RWI has been previously proposed to explain several phenomena occurring in the vicinity of black-holes
      \citep{Vin13,Vin14} but its existence in a full-GR environment had never been demonstrated up to now.
            Here we remedy this by performing 2D and 3D GR hydrodynamics simulations of the RWI that enable us to prove at the same time: \\
\noindent i) that the instability actually exists when taking into account the full effects of the non-spinning black hole
	
\noindent ii) and that its local behaviour is only mildly affected by the strong gravity generated from a non-spinning black hole.

    Nevertheless, while the GR effects are mild on the growth and saturation of the instability, this is not what an observer at infinity
    would perceive first because the local clock rate depends on the location of the emitting region but also because of the way the emitted photons would
     travel to the observer.  
    Indeed, a remote observer will perceive distorted instability patterns as a consequence of the time dilatation occurring near the black hole.
    In the case of a Schwarzschild black-hole  as presented here, this effect is 
     relatively weak as the typical variation length scale of the lapse function $\alpha$  is large compared to the size of the region where the instability develops. 
    This implies that in order to depict the RWI near non-spinning black hole, we can keep a mostly pseudo-Newtonian approach for the fluid dynamics as long 
    as we take into account the time dilation. 
    The second effect needed to compute observables is to take into account the way the emitted light would arrive to the 
    observer through GR raytracing  \citep[see e.g.][]{Fala07}.  
    In the case of a Schwarzschild black hole this will be limited to high inclination systems and still would be hard to constrain as it is entangled with other 
    effects for most of the observables we have access to, see for example \citet{VV16} for the impact on an instability of the same family.
    In order to see some potential effect in observations one would need enough precision in time to follow the changes in shape of the pulse profile within 
    a few orbits. Indeed, while the mode structure is not widely influenced by GR, the way it is perceived by an observer changes as seen in \citet{VV16} 
    and can lead to detectable differences in the pulse profile. Adressing such differences, in the context of a more realistic setup for observables like the
    flares of Sgr A$^\star$, is beyond the scope of this paper as it would also require to take into account the potential spin of the black hole.

\section*{Acknowledgments}
 We acknowledge the financial support from the UnivEarthS Labex program of Sorbonne Paris Cit\'e (ANR-10-LABX-0023 and ANR-11-IDEX-0005-02). 

\bibliographystyle{mn2e}
\bibliography{biblio}

\begin{thebibliography}{}
\makeatletter
\relax
\def\mn@urlcharsother{\let\do\@makeother \do\$\do\&\do\#\do\^\do\_\do\%\do\~}
\def\mn@doi{\begingroup\mn@urlcharsother \@ifnextchar [ {\mn@doi@}
  {\mn@doi@[]}}
\def\mn@doi@[#1]#2{\def\@tempa{#1}\ifx\@tempa\@empty \href
  {http://dx.doi.org/#2} {doi:#2}\else \href {http://dx.doi.org/#2} {#1}\fi
  \endgroup}
\def\mn@eprint#1#2{\mn@eprint@#1:#2::\@nil}
\def\mn@eprint@arXiv#1{\href {http://arxiv.org/abs/#1} {{\tt arXiv:#1}}}
\def\mn@eprint@dblp#1{\href {http://dblp.uni-trier.de/rec/bibtex/#1.xml}
  {dblp:#1}}
\def\mn@eprint@#1:#2:#3:#4\@nil{\def\@tempa {#1}\def\@tempb {#2}\def\@tempc
  {#3}\ifx \@tempc \@empty \let \@tempc \@tempb \let \@tempb \@tempa \fi \ifx
  \@tempb \@empty \def\@tempb {arXiv}\fi \@ifundefined
  {mn@eprint@\@tempb}{\@tempb:\@tempc}{\expandafter \expandafter \csname
  mn@eprint@\@tempb\endcsname \expandafter{\@tempc}}}

\bibitem[\protect\citeauthoryear{{Falanga}, {Melia}, {Tagger}, {Goldwurm}  \&
  {B{\'e}langer}}{{Falanga} et~al.}{2007}]{Fala07}
{Falanga} M.,  {Melia} F.,  {Tagger} M.,  {Goldwurm} A.,   {B{\'e}langer} G.,
  2007, \mn@doi [\apjl] {10.1086/519278}, \href
  {http://cdsads.u-strasbg.fr/abs/2007ApJ...662L..15F} {662, L15}

\bibitem[\protect\citeauthoryear{{Hawley}, {Smarr}  \& {Wilson}}{{Hawley}
  et~al.}{1984}]{Hawley84}
{Hawley} J.~F.,  {Smarr} L.~L.,   {Wilson} J.~R.,  1984, \mn@doi [\apj]
  {10.1086/161696}, \href {http://cdsads.u-strasbg.fr/abs/1984ApJ...277..296H}
  {277, 296}

\bibitem[\protect\citeauthoryear{{Koren}}{{Koren}}{1993}]{Koren93}
{Koren} B.,  1993, Numerical Methods for Advection Diffusion Problems,
  Braunschweig: Vieweg, p. 117, 1, 117

\bibitem[\protect\citeauthoryear{{Li}, {Finn}, {Lovelace}  \& {Colgate}}{{Li}
  et~al.}{2000}]{Li00}
{Li} H.,  {Finn} J.~M.,  {Lovelace} R.~V.~E.,   {Colgate} S.~A.,  2000, \mn@doi
  [\apj] {10.1086/308693}, \href
  {http://cdsads.u-strasbg.fr/abs/2000ApJ...533.1023L} {533, 1023}

\bibitem[\protect\citeauthoryear{{Li}, {Colgate}, {Wendroff}  \& {Liska}}{{Li}
  et~al.}{2001}]{Li01}
{Li} H.,  {Colgate} S.~A.,  {Wendroff} B.,   {Liska} R.,  2001, \mn@doi [\apj]
  {10.1086/320241}, \href {http://cdsads.u-strasbg.fr/abs/2001ApJ...551..874L}
  {551, 874}

\bibitem[\protect\citeauthoryear{{Lin}}{{Lin}}{2012}]{Lin12}
{Lin} M.-K.,  2012, \mn@doi [\mnras] {10.1111/j.1365-2966.2012.21955.x}, \href
  {http://cdsads.u-strasbg.fr/abs/2012MNRAS.426.3211L} {426, 3211}

\bibitem[\protect\citeauthoryear{{Lovelace} \& {Hohlfeld}}{{Lovelace} \&
  {Hohlfeld}}{1978}]{Lovelace78}
{Lovelace} R.~V.~E.,  {Hohlfeld} R.~G.,  1978, \mn@doi [\apj] {10.1086/156004},
  \href {http://cdsads.u-strasbg.fr/abs/1978ApJ...221...51L} {221, 51}

\bibitem[\protect\citeauthoryear{{Lovelace} \& {Romanova}}{{Lovelace} \&
  {Romanova}}{2014}]{Lovelace14}
{Lovelace} R.~V.~E.,  {Romanova} M.~M.,  2014, \mn@doi [Fluid Dynamics
  Research] {10.1088/0169-5983/46/4/041401}, \href
  {http://cdsads.u-strasbg.fr/abs/2014FlDyR..46d1401L} {46, 041401}

\bibitem[\protect\citeauthoryear{{Lovelace}, {Li}, {Colgate}  \&
  {Nelson}}{{Lovelace} et~al.}{1999}]{Lovelace99}
{Lovelace} R.~V.~E.,  {Li} H.,  {Colgate} S.~A.,   {Nelson} A.~F.,  1999,
  \mn@doi [\apj] {10.1086/306900}, \href
  {http://cdsads.u-strasbg.fr/abs/1999ApJ...513..805L} {513, 805}

\bibitem[\protect\citeauthoryear{{Lyra} \& {Mac Low}}{{Lyra} \& {Mac
  Low}}{2012}]{Lyra12}
{Lyra} W.,  {Mac Low} M.-M.,  2012, \mn@doi [\apj]
  {10.1088/0004-637X/756/1/62}, \href
  {http://cdsads.u-strasbg.fr/abs/2012ApJ...756...62L} {756, 62}

\bibitem[\protect\citeauthoryear{{Mathews}}{{Mathews}}{1971}]{Mat71}
{Mathews} W.~G.,  1971, \mn@doi [\apj] {10.1086/150883}, \href
  {http://cdsads.u-strasbg.fr/abs/1971ApJ...165..147M} {165, 147}

\bibitem[\protect\citeauthoryear{{Meheut}, {Casse}, {Varniere}  \&
  {Tagger}}{{Meheut} et~al.}{2010}]{Meheut10}
{Meheut} H.,  {Casse} F.,  {Varniere} P.,   {Tagger} M.,  2010, \mn@doi [\aap]
  {10.1051/0004-6361/201014000}, \href
  {http://cdsads.u-strasbg.fr/abs/2010A%26A...516A..31M} {516, A31}

\bibitem[\protect\citeauthoryear{{Meliani}, {Sauty}, {Tsinganos}  \&
  {Vlahakis}}{{Meliani} et~al.}{2004}]{Meliani04}
{Meliani} Z.,  {Sauty} C.,  {Tsinganos} K.,   {Vlahakis} N.,  2004, \mn@doi
  [\aap] {10.1051/0004-6361:20035653}, \href
  {http://cdsads.u-strasbg.fr/abs/2004A%26A...425..773M} {425, 773}

\bibitem[\protect\citeauthoryear{{Mignone} \& {McKinney}}{{Mignone} \&
  {McKinney}}{2007}]{Mignone07}
{Mignone} A.,  {McKinney} J.~C.,  2007, \mn@doi [\mnras]
  {10.1111/j.1365-2966.2007.11849.x}, \href
  {http://cdsads.u-strasbg.fr/abs/2007MNRAS.378.1118M} {378, 1118}

\bibitem[\protect\citeauthoryear{{Porth}, {Xia}, {Hendrix}, {Moschou}  \&
  {Keppens}}{{Porth} et~al.}{2014}]{Porth14}
{Porth} O.,  {Xia} C.,  {Hendrix} T.,  {Moschou} S.~P.,   {Keppens} R.,  2014,
  \mn@doi [ApJS] {10.1088/0067-0049/214/1/4}, \href
  {http://cdsads.u-strasbg.fr/abs/2014ApJS..214....4P} {214, 4}

\bibitem[\protect\citeauthoryear{{Tagger} \& {Melia}}{{Tagger} \&
  {Melia}}{2006}]{Tagger06}
{Tagger} M.,  {Melia} F.,  2006, \mn@doi [\apjl] {10.1086/499806}, \href
  {http://cdsads.u-strasbg.fr/abs/2006ApJ...636L..33T} {636, L33}

\bibitem[\protect\citeauthoryear{{Tagger} \& {Varni{\`e}re}}{{Tagger} \&
  {Varni{\`e}re}}{2006}]{TV06}
{Tagger} M.,  {Varni{\`e}re} P.,  2006, \mn@doi [\apj] {10.1086/508318}, \href
  {http://cdsads.u-strasbg.fr/abs/2006ApJ...652.1457T} {652, 1457}

\bibitem[\protect\citeauthoryear{{Taub}}{{Taub}}{1948}]{Taub48}
{Taub} A.~H.,  1948, \mn@doi [Physical Review] {10.1103/PhysRev.74.328}, \href
  {http://cdsads.u-strasbg.fr/abs/1948PhRv...74..328T} {74, 328}

\bibitem[\protect\citeauthoryear{{Varni{\`e}re} \& {Tagger}}{{Varni{\`e}re} \&
  {Tagger}}{2006}]{VarnT06}
{Varni{\`e}re} P.,  {Tagger} M.,  2006, \mn@doi [\aap]
  {10.1051/0004-6361:200500226}, \href
  {http://cdsads.u-strasbg.fr/abs/2006A%26A...446L..13V} {446, L13}

\bibitem[\protect\citeauthoryear{{Varniere} \& {Vincent}}{{Varniere} \&
  {Vincent}}{2016}]{VV16}
{Varniere} P.,  {Vincent} F.~H.,  2016, \mn@doi [\aap]
  {10.1051/0004-6361/201527711}, \href
  {http://cdsads.u-strasbg.fr/abs/2016A%26A...591A..36V} {591, A36}

\bibitem[\protect\citeauthoryear{{Varniere}, {Tagger}  \&
  {Rodriguez}}{{Varniere} et~al.}{2011}]{VTR11}
{Varniere} P.,  {Tagger} M.,   {Rodriguez} J.,  2011, \mn@doi [\aap]
  {10.1051/0004-6361/201015028}, \href
  {http://cdsads.u-strasbg.fr/abs/2011A%26A...525A..87V} {525, A87}

\bibitem[\protect\citeauthoryear{{Varni{\`e}re}, {Tagger}  \&
  {Rodriguez}}{{Varni{\`e}re} et~al.}{2012}]{VTR12}
{Varni{\`e}re} P.,  {Tagger} M.,   {Rodriguez} J.,  2012, \mn@doi [\aap]
  {10.1051/0004-6361/201116698}, \href
  {http://cdsads.u-strasbg.fr/abs/2012A%26A...545A..40V} {545, A40}

\bibitem[\protect\citeauthoryear{{Vincent}, {Meheut}, {Varniere}  \&
  {Paumard}}{{Vincent} et~al.}{2013}]{Vin13}
{Vincent} F.~H.,  {Meheut} H.,  {Varniere} P.,   {Paumard} T.,  2013, \mn@doi
  [\aap] {10.1051/0004-6361/201220695}, \href
  {http://cdsads.u-strasbg.fr/abs/2013A%26A...551A..54V} {551, A54}

\bibitem[\protect\citeauthoryear{{Vincent}, {Paumard}, {Perrin}, {Varniere},
  {Casse}, {Eisenhauer}, {Gillessen}  \& {Armitage}}{{Vincent}
  et~al.}{2014}]{Vin14}
{Vincent} F.~H.,  {Paumard} T.,  {Perrin} G.,  {Varniere} P.,  {Casse} F.,
  {Eisenhauer} F.,  {Gillessen} S.,   {Armitage} P.~J.,  2014, \mn@doi [\mnras]
  {10.1093/mnras/stu812}, \href
  {http://cdsads.u-strasbg.fr/abs/2014MNRAS.441.3477V} {441, 3477}

\bibitem[\protect\citeauthoryear{{van der Holst}, {Keppens}, {Meliani},
  {Porth}, {van Marle}, {Delmont}  \& {Xia}}{{van der Holst}
  et~al.}{2012}]{Holst12}
{van der Holst} B.,  {Keppens} R.,  {Meliani} Z.,  {Porth} O.,  {van Marle}
  A.~J.,  {Delmont} P.,   {Xia} C.,  2012, {MPI-AMRVAC: MPI-Adaptive Mesh
  Refinement-Versatile Advection Code}, Astrophysics Source Code Library
  (\mn@eprint {ascl} {1208.014})

\makeatother
\end{thebibliography}

\end{document}